\documentclass[prd,floats,superscriptaddress,nofootinbib,preprintnumbers,onecolumn]{revtex4}%
\usepackage[normalem]{ulem}
\usepackage{lipsum, babel}
\usepackage[a4paper, margin=2cm]{geometry}
\usepackage{amssymb}
\usepackage{amsmath}

\usepackage{sans}

\makeatletter\AtBeginDocument{\let\@elt\relax}\makeatother

\usepackage{graphicx}
\usepackage{graphics}
\usepackage{dcolumn}
\usepackage{color}
\usepackage{rotate}
\usepackage{fancyhdr} 
\usepackage{hyperref}
\usepackage{indentfirst}
\usepackage{booktabs} 
\usepackage{fontawesome} 
\usepackage{physics} 
\usepackage{siunitx} 
\DeclareSIUnit \parsec {pc}
\DeclareSIUnit \year {yr}
\usepackage{umoline}
\usepackage{cancel}
\usepackage{tikz}
\usetikzlibrary{arrows.meta,decorations.markings,calc}

\def\be{\begin{eqnarray}}
\def\ee{\end{eqnarray}}
\newcommand{\sgn}{\mathrm{Sgn}}
\newcommand{\iu}{\mathrm{i}}
\newcommand{\e}{\mathrm{e}}

\begin{document}
\title{Gravitational lensing time delay beyond the Shapiro/geometry split}
 \author{Luca Teodori}\email{luca.teodori@iac.es}  
 \affiliation{Instituto de Astrof\'isica de Canarias, C/ V\'ia L\'actea, s/n E38205, La Laguna, Tenerife, Spain
 	} 
 \affiliation{Universidad de La Laguna, Departamento de Astrof\'isica, La Laguna, Tenerife,
 	Spain}
\author{Kfir Blum}\email{kfir.blum@weizmann.ac.il}
\affiliation{Weizmann Institute of Science, Rehovot 7610001, Israel} 
 \author{Zhaoyu Bai}\email{zhaoyu.bai@weizmann.ac.il}  
\affiliation{Weizmann Institute of Science, Rehovot 7610001, Israel} 

\begin{abstract}
Time delays are a key observable in strong gravitational lensing systems. Their theoretical expression is usually written as a sum of a geometrical delay and a Shapiro delay, with cosmology entering through angular diameter distances and a redshift prefactor. In this work we derive this structure from the exact null geodesics of the Schwarzschild-de Sitter metric. The standard formula is recovered as the leading term in a small-angle expansion, and we identify the first correction to the usual geometrical-plus-Shapiro split. Such correction does not introduce any new cosmological dependence: it corresponds instead to a higher-order correction intrinsic to the Schwarzschild part of the metric. As a consequence, up to this order, the cosmological constant enters only through the unlensed angular diameter distances and the unlensed lens-redshift prefactor.
\end{abstract}

\maketitle
\section{Introduction} \label{s:intro}
One key prediction of Einstein's general relativity is the deflection of light by intervening matter, which acts as a gravitational lens between an emitter and an observer. Important examples of gravitational lensing happen on cosmological scales, where the lens is a massive galaxy or cluster and the emitters are supernovae or quasars at Gpc distances. In such systems there is a large separation of scales between cosmological distances and the physical extent of the lens. This separation motivates the standard treatment in which the cosmological background and the local lens contribution are handled separately. In practice, in the thin-lens and weak-field approximations, the light deflection is assumed to happen only in the vicinity of the lens, where cosmological expansion can be neglected~\cite{Schneider:1992,Seitz:1994xf,Bartelmann:2010fz}. Cosmology then enters the gravitational lensing observables through angular diameter distances. 

For most current strong-lensing applications, including quasar time-delay cosmography~\cite{Suyu:2016qxx,Millon:2019slk}, this separation is an excellent approximation. Precision lensing measurements, however, motivate a more explicit understanding of the expansion behind it~\cite{Sasaki:1993tu,Wucknitz:2020spz}. In particular, it is useful to ask which dimensionless parameter controls the usual formulae and what the first correction to them is.

A useful exact setting for this question is the Schwarzschild-de Sitter (SdS) metric. The SdS metric (see e.g.~\cite{Lake:1977ui,Perlick:2004tq}) is an exact solution of the Einstein equations with a point mass, parametrized by the Schwarzschild radius $ r_{\rm s} $, and a positive cosmological constant $ \Lambda $. More generally, the same static point-mass solution with arbitrary sign of $ \Lambda $ is often called the Kottler metric~\cite{1918AnP...361..401K}. The negative-$\Lambda$ case is Schwarzschild-anti de Sitter. In this work we restrict to $ \Lambda>0 $. The SdS metric is clearly idealized: realistic lensing systems are not exactly de Sitter and their lenses are not point masses. Its advantage is that the null geodesics are sufficiently explicit to let us compare exact expressions with the standard lensing expansion. This metric has been studied extensively in connection with the deflection angle and the question of whether $ \Lambda $ has an independent lensing effect~\cite{Piattella:2016nzt,Rindler:2007zz,Ishak:2010zh,Aghili:2014aga,Hu:2021yzn,Sereno:2008kk,Arakida:2011ty}, beyond its effect on angular diameter distances and redshift expressions. We find no independent $ \Lambda $ contribution to the lens equation through the order relevant here, in agreement with Ref.~\cite{Butcher:2016yrs}.

Time delays provide a second central strong-lensing observable~\cite{1975ApJ...195L..11C}, used both for lens modeling and for cosmology~\cite{Refsdal:1964nw,Millon:2019slk}. The role of $ \Lambda $ in time-delay calculations has received less systematic attention than in deflection-angle calculations. Earlier work includes discussions of Shapiro time delay in SdS for solar-system settings~\cite{Kagramanova:2006ax}, applications to galaxy lensing in SdS~\cite{Schucker:2008jv}, and studies in related McVittie or Einstein-Straus geometries~\cite{McVittie:1933zz,Kaloper:2010ec,Park:2008ih,Piattella:2016nzt,Bessa:2022sdh,Arakida:2011st,Einstein:1945id,1954ZPhy..137..595S,Schucker:2008bc,Boudjemaa:2011ab,Chen:2010gi} (see also~\cite{Alchera:2017sjt,Alchera:2018ijo} for a different approach to get corrections to the standard time delay formula). Here we keep the simple SdS setting.

Our goal is to derive the standard lens equation and time-delay formula as the leading terms of the exact SdS geodesic expressions. We work with a comoving emitter and observer in static coordinates and expand in the small parameter $x=r_{\rm s}/r_{\rm c}\sim\theta$, where $r_{\rm c}$ is the closest-approach radius of the light ray. We show how the familiar division of the time delay into a geometrical term and a Shapiro term appears after the small-angle and large-distance expansion, and after rewriting the answer in terms of image positions and angular diameter distances, and identify the first correction to this split. This correction does not introduce independent $\Lambda$ contributions to the time-delay expression, and it can be traced back to the Schwarzschild part of the metric. 

This paper is structured as follows. In Sec.~\ref{s:light_time_delay} we collect the exact SdS equations for light deflection and time delays. In Sec.~\ref{s:small_angle} we derive the small-angle, large-distance expansion and obtain the leading time-delay formula together with its first correction. We conclude in Sec.~\ref{s:summary}. The appendices collect supporting material on peculiar velocities, the FLRW redshift limit, angular diameter distances, astrophysical values of the expansion parameter, and miscellaneous technical aspect of some calculations, respectively.

\section{Light deflection and time delays in the Schwarzschild-de Sitter metric} \label{s:light_time_delay}
In this section we collect the exact expressions for light deflection and time delays in the SdS metric. 

\subsection{Geodesic equations} \label{s:geodesics}
The SdS metric in static coordinates  reads
\be \label{eq:metric}
	\dd{s^2} = -A(r) \dd{t^2} + A^{-1}(r) \dd{r^2}  + r^2\dd{\Omega^2} \ , \ A(r) := 1 - \frac{r_{\rm s}}{r} - H^2r^2 \ , \ \dd{\Omega} := \dd{\phi}^2 + \sin^2\phi \dd{\varphi}^2 \ ,
\ee
where $ r_{\rm s} $ is the Schwarzschild radius and $ H $ is the Hubble constant (connected to the cosmological constant $ \Lambda $ via $ H^2 = \Lambda/3  $). We focus on the case $ \Lambda > 0 $.
Horizons arise at $r=r_{\rm h}$ defined by the solutions of
\be
	A(r_{\rm h}) = 0 \implies  H^2 \frac{r^3_{\rm h}}{r_{\rm s}} - \frac{r_{\rm h}}{r_{\rm s}} +1 = 0  \ .
\ee
In the regime $ r_{\rm s} H \ll 1 $, the two positive real solutions $ r_{\rm h} \approx r_{\rm s} $ and $ r_{\rm h} \approx 1/H $ are
the point mass event horizon and the cosmological horizon respectively. The static coordinates cover the space between these two horizons.

Exploiting spherical symmetry, we set the geodesic motion in the equatorial plane, $\phi=\pi/2$. With $ \ \dot{ } := \dv*{\lambda} $, where $ \lambda  $ is the affine parameter, the relevant first integrals are
\be\label{dott}
	\dot{t} = \frac{E}{A} \ ,
\ee
\be\label{dotvarphi}
	\dot{\varphi} = \frac{b}{r^2} \ ,
\ee
\be\label{rdot}
	\dot{r} = \pm \sqrt{Ae + E^2 - \frac{A b^2}{r^2}} \ . 
\ee
Here $E$ and $b$ are the conserved quantities associated with the $t$ and $\varphi$ Killing directions, and $g_{\mu\nu}\dot{x}^\mu \dot{x}^\nu=e$, with $e=0$ for null geodesics and $e<0$ for timelike geodesics. The sign in Eq.~\eqref{rdot} should be chosen according to whether the geodesic has decreasing or increasing $ r $. We normalize the massive geodesic so that $ e=-1 $. With such a choice, $ \lambda $ will be the massive geodesic proper time.

Notice that it is possible to rescale away $ E $ by redefining the affine parameter $ \lambda $ in the massless case. Hence, we will always set $ E =1 $ for massless geodesics. This is not possible in the massive case. However, for comoving observers in the de Sitter expansion, $ E=1 $ is a natural choice, as explained in Ref.~\cite{Butcher:2016yrs}. In the main text we set $E_{\rm e}=E_{\rm o}=1$ for the emitter and observer. Deviations from this choice can be interpreted as peculiar velocities of the massive geodesic. For completeness, we summarize the corresponding bookkeeping in App.~\ref{s:peculiar}.

\subsection{Lensing set-up} \label{s:lensing_setup}
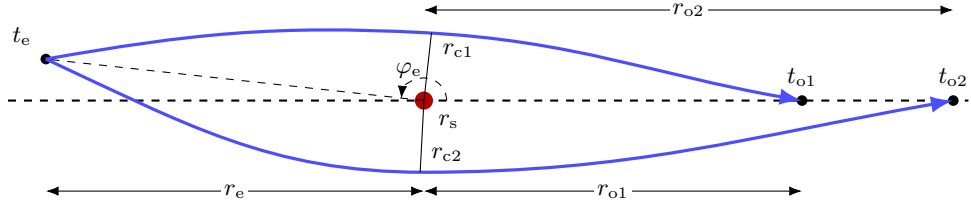
\begin{figure}
	\centering
	\begin{tikzpicture}[x=1cm,y=1cm,>=Latex,thick]
		\def\xS{-2.0}
		\def\xL{3.0}
		\def\xO{8.0}
		\def\xOtwo{10.0}
		\def\yS{0.55}
		\def\b{0.90}
		\def\btwo{0.95}
		\def\massR{0.12}
		\draw[dashed] (\xS-0.5,0) -- (\xOtwo+0.2,0) node[right] {};
		\fill (\xS,\yS) circle (2pt) node[above left=2pt] {$t_{\rm e}$};
		\fill[red!70!black] (\xL,0) circle (\massR);
		\node[below right=2pt] at (\xL,0) {$r_{\rm s}$};
		\fill (\xO,0) circle (2pt) node[above=2pt] {$t_{\rm o1}$};
		\fill (\xOtwo,0) circle (2pt) node[above=2pt] {$t_{\rm o2}$};
		\path
		(\xS,\yS) coordinate (S)
		(\xL, \b) coordinate (Lup)
		(\xO, 0) coordinate (Ou)
		(\xOtwo, 0) coordinate (Ol);
		\path
		(\xL,-\btwo) coordinate (Ldown);
		\draw[blue!70, line width=1.2pt, ->]
		(S) to[out=10, in=177] (Lup)
		to[out=-3, in=170] (Ou);
		\draw[blue!70, line width=1.2pt, ->]
		(S) to[out=-25, in=180] (Ldown)
		to[out=0, in=190] (Ol);
		\draw[<->,thin] (\xS,-1.2) -- node[fill=white,inner sep=1pt] {$r_{\rm e}$} (\xL,-1.2);
		\draw[<->,thin] (\xL,-1.2) -- node[fill=white,inner sep=1pt] {$r_{\rm o1}$} (\xO,-1.2);
		\draw[<->,thin] (\xL,1.2) -- node[fill=white,inner sep=1pt] {$r_{\rm o2}$} (\xOtwo,1.2);
		\draw[thin] (\xL,0) -- (\xL+0.1,0.9) node[below right=1pt] {$r_{\rm c1}$};
		\draw[thin] (\xL,0) -- (\xL-0.05,-0.95) node[above right=1pt] {$r_{\rm c2}$};
		\draw[dashed, thin] (\xL,0) -- (\xS,\yS);
		\draw[->, dashed, thin] (\xL+0.3,0) arc[start angle=0, end angle=175, radius=0.3];
		\node at (\xL-0.2,0.4) {$\varphi_{\rm e}$};
	\end{tikzpicture}
	\caption{Illustration of the lensing setup in SdS static coordinates.}
	\label{fig:setup}
\end{figure}

We now describe gravitational lensing in this metric, where light rays coming from a single emission event by a comoving emitter get deflected by the point mass at the origin, travel along two separate geodesics, and reach a comoving observer at separate observation events. We remind that we consider comoving observers as observers characterized by a 4-velocity with $E=1$, see Eq.~\ref{dott}.
We denote by $ \omega_i^\mu $ ($i=1,2$) the tangent vector of the light ray propagating along the $i$'th null geodesic from the emission event $ x_{\rm e} = (t_{\rm e}, r_{\rm e}, \pi/2, \varphi_{\rm e}) $ to the $i$'th observation event $x_{\mathrm{o}i} = (t_{\mathrm{o}i}, r_{\mathrm{o}i}, \pi/2, \varphi_{\rm o}) $. We are interested in small deflections, wherein $ \varphi_{\rm e} \sim \pi > \varphi_{\rm o} \sim 0 $. The setup is illustrated in Fig.~\ref{fig:setup}. 

Let $ u_{\rm e} $ be the 4-velocity of the emitter and $ u_{\rm o} $ the 4-velocity of the observer. We assume that both emitter and observer move along lines with constant $ \varphi $. 
Notice that $  r_{\mathrm{o}i} $ differs for different light rays due to the static coordinates used\footnote{In particular, if de Sitter expansion dominates over the point-mass attraction for the observer under consideration, one expects $r_{\rm o1} < r_{\rm o2}$. }. 
We have 
\be \label{eq:omega_mu}
	\omega^\mu_i = \qty(\frac{1}{A(r)}, \pm\sqrt{1-\frac{b_i^2 A(r)}{r^2}},0, \frac{b_i}{r^2} ) \ ,
\ee
\be \label{eq:u_mu}
	\begin{aligned}
	u^\mu_{\rm e} &= \qty(\frac{E_{\rm e}}{A(r)}, \sqrt{ E^2_{\rm e}-A(r)},0, 0 ) \ , \\
	u^\mu_{\rm o} &= \qty(\frac{E_{\rm o}}{A(r)}, \sqrt{ E^2_{\rm o}-A(r)},0, 0 ) \ .
	\end{aligned}
\ee
where we kept explicit the factors $E_{\rm o}, E_{\rm e}$ for now.
The sign on $ \dot{r} $ is positive for the massive geodesics since we assume the set-up to be such that the emitter and observer's motion are dominated by the Hubble expansion, rather than the point mass. The sign for the light geodesics is chosen according to whether the light ray is before (minus) or after (plus) the closest approach to the point mass. We define $ b_i =: s_i |b_i| $, where $ s_i = -1 $ if the light moves on the orbit plane in a clockwise direction, and $ s_i=1 $  otherwise. 

\subsection{Deflection angle and image positions} \label{s:deflection_image_positions}
The deflection angle of geodesic $\omega_i^\mu$ is given by
\be\label{b_i}
	\begin{aligned}
		&\int_{r_{\rm e}}^{r_{\mathrm{o} i}} \qty(\dv{\varphi}{r})_{\omega_i^\mu} \dd{r} =  \varphi_{\rm o} - \varphi_{\rm e} + \pi(1+s_i) =  -\int_{r_{\rm e}}^{r_{\mathrm{c} i}} \frac{b_i\dd{r}}{r\sqrt{r^2 -b_i^2A(r)}} +\int_{r_{\mathrm{c}i}}^{r_{\mathrm{o}i}} \frac{b_i\dd{r}}{r\sqrt{r^2 -b_i^2A(r)}} \\ 
		&= s_i\qty( \int_{u_{\rm e}}^{u_{\mathrm{c}i}}  + \int_{u_{\mathrm{o}i}}^{u_{\mathrm{c}i}} )  \frac{\dd{u}}{\sqrt{\frac{r^2_{\rm s}}{\tilde{b}_i^2} -u^2(1-u) }} \ .
	\end{aligned}
\ee
This is identical to the expression in a pure Schwarzschild metric, apart from the re-definition $ b_i \to \tilde{b}_i $. 
We defined the dimensionless inverse radius 
\be
u := r_{\rm s}/r,
\ee
so that, e.g., $u_{\rm e}=r_{\rm s}/r_{\rm e}$, etc. We use this notation below. 
The closest approach distance $ r_{\mathrm{c}i} $ of the $i$'th light geodesic to the central point mass is given by a positive root of the cubic equation:
\be \label{rc}  
	r^3 - r\tilde{b}_i^2 + \tilde{b}_i^2r_{\rm s} = 0 \ , \ 
	\tilde{b}_i = \frac{b_i}{\sqrt{1+H^2b_i^2}} \ . 
\ee
In terms of $u$, this equation reads
\be\label{eq:uceq}
	u^2(1-u)\ = \ \frac{r^2_{\rm s}}{\tilde{b}^2_i} \ .
\ee
For $r_{\rm s}/|\tilde b_i|\ll1$, the turning point solution $u_{{\rm c}i}$ satisfies $u_{{\rm c}i}=r_{\rm s}/|\tilde b_i|+r_{\rm s}^2/(2\tilde b_i^2)+5r_{\rm s}^3/(8|\tilde b_i|^3)+...$\, .

Image position angles are usually defined with respect to a light ray directed along $ \varphi = \varphi_{\rm o} $, pointing from the lens to the observer. To such light ray we assign the geodesic vector
\be \label{eq:nmu}
	n^\mu = \qty(\frac{1}{A(r)}, 1,0,0 ) \ . 
\ee
The observable image position angle $ \theta_i $ formed from the light ray $ n^\mu $ and $ \omega_i^\mu $ with respect to the observer $ u^\mu_{{\rm o}} $ at $r_{{\rm o}i}$ reads
\be \label{angle_spec}
	1 - \cos\theta_i = - \eval{\frac{g_{\mu\nu}\omega_i^\mu n^\nu }{(g_{\mu\nu} u_{\rm o}^\mu n^\nu)(g_{\mu\nu} u_{\rm o}^\mu \omega_i^\nu) }}_{x_{\mathrm{o}i}} = \frac{(1- \sqrt{1 - b_i^2A(r_{\mathrm{o}i})/r_{\mathrm{o}i}^2}) (E_{\rm o}+\sqrt{E^2_{\rm o}-A(r_{\mathrm{o}i})  }) }{ E_{\rm o}-\sqrt{1-b_i^2A(r_{\mathrm{o}i}) /r_{\mathrm{o}i}^2 } \sqrt{E^2_{\rm o}-A(r_{\mathrm{o}i})}} \ .
\ee

In the main text below, Eq.~\eqref{angle_spec} will be evaluated for $E_{\rm o}=1$. The more general expression is useful if one wants to add observer peculiar velocity effects. See App.~\ref{s:peculiar}.

\subsection{Redshift} \label{s:exact_redshift}
The redshifts associated with the lens and with the emitter images as seen by the observer are also measurable. 
The redshift associated with a light ray $ \omega^\mu_i $ coming from an emitter $ u^\mu_{\rm e} $ at $ x_{\rm e} $ as seen by an observer $ u^\mu_{\rm o} $ at $ x_{\mathrm{o}i} $ reads
\be \label{zs}
	1 + z_{\mathrm{e}i} :=\frac{g_{\mu\nu}(x_{\rm e})\omega^\mu(x_{\rm e}) u^\nu_{\rm e}(x_{\rm e})}{g_{\mu\nu}(x_{\mathrm{o}i})\omega^\mu(x_{\mathrm{o}i}) u^\nu_{\rm o}(x_{\mathrm{o}i})}  = \frac{A(r_{\mathrm{o}i}) }{A(r_{\rm e})} \qty( \frac{E_{\rm e} + \sqrt{E^2_{\rm e} - A(r_{\rm e})} 
		\sqrt{1-b_i^2A(r_{\rm e})/r^2_{\rm e}}}{E_{\rm o} - \sqrt{E^2_{\rm o} - A(r_{\mathrm{o}i})} 
		\sqrt{1-b_i^2A(r_{\mathrm{o}i})/r^2_{\mathrm{o}i}}} ) \ .
\ee
We can also write an expression for the lens redshift, to be computed using a light ray originating near the lens. 
For simplicity, we can assume that the light ray used to measure the lens redshift is emitted with $ b=0 $, from an emitter sitting at $ r_{\rm lens} $ with 4-velocity 
\be \label{eq:generic_obs}
	u^\mu_{\rm l} = \qty(\frac{E_{\rm l}}{A}, \sqrt{E_{\rm l}^2 - A - A\frac{b^2_{\rm l}}{r^2}} ,0, \frac{b_{\rm l}}{r^2})  \ .
\ee
Here $ b_{\rm l} $ parameterizes $ u^\mu_{\rm l} $, and should not be confused with the light ray geodesic which has $ b=0 $. A source of lens light located near the lens need not be comoving with the de Sitter flow. In the present idealized setup, one possible lens-redshift expression reads
\be \label{eq:zl_alpha}
	1 + z_{\rm l} = \frac{E_{\rm o}+\sqrt{E^2_{\rm o}-A(r_{\rm o})}}{A(r_{\rm lens})} (E_{\rm l}-J(r_{\rm lens})) \ , \  J(r) :=\sqrt{E_{\rm l}^2- A(r)\qty(1+ \frac{b^2_{\rm l}}{r^2}) } \ .
\ee

  The spectroscopic redshift of light emitted near the lens,   $z_{\rm l}$, includes local contributions from the emitting material and from the observer prescription used to define the lens distance. The standard lensing notation uses the background lens-plane redshift associated with the observer-lens angular-diameter-distance convention. We denote this quantity by   $\tilde z_{\rm l}$. In the SdS calculation, $\tilde z_{\rm l}$ is the natural parameter that trades the static-coordinate radius $r_{\rm o}$ for the unlensed observer-lens distance $D_{\rm l}$,
 defined below in Eq.~\eqref{eq:Dd_main}, without including the local $r_{\rm lens}\neq0$ and $E_{\rm l}\neq1$ details of Eq.~\eqref{eq:zl_alpha}. At leading order it agrees with the usual redshift-distance relation: in the $r_{\rm s}=0$ limit,
\be \label{eq:redshift_r}
	H r_{\rm o} = z_{\rm l} \ , \ H r_{\rm e} = 1 - \frac{1 + z_{\rm l}}{1+z_{\rm e}} \ ,
\ee
as shown in App.~\ref{s:comoving}. As mentioned above, measured lens redshift also depends on local lens physics such as gravitational potentials, velocity dispersion, rotation, aperture weighting, and spectral modeling. These effects are outside the point-mass SdS idealization and should not be identified with a universal cosmological correction.

\subsection{Time delay} \label{s:exact_time_delay}
We are interested in the time delay between image $ 1 $ and $ 2 $ of two light rays emitted from the same event $ x_{\rm e} $, and take image $1$ to arrive before image $2$.
Denoting the proper time of the observer as $ T $ (the time indicated on their wrist watch), we can use
\be \label{DeltaT}
	\dot{r} =: \dv{r}{T} \implies \Delta T := T_{\mathrm{o}2 } - T_{\mathrm{o}1 }  = \int_{r_{\mathrm{o}1}}^{r_{\mathrm{o}2}} \frac{\dd{r}}{\sqrt{E^2_{\rm o}-1 + r_{\rm s}/r + H^2 r^2}} \ .
\ee
%
%
The light-ray geodesics give the same static coordinate time difference\footnote{Note that both Eq.~\eqref{b_i} and Eq.~\eqref{eq:time_delay} are consistent with the deflection angle and time delay formulae for a generic static spherically symmetric metric~\cite{Perlick:2003vg,Perlick:2015ncm}.}:
\be\label{eq:time_delay}
	\begin{aligned}
		&t_{\rm o2} - t_{\rm o1}=\int_{r_{\rm e}}^{r_{\rm o2}} \qty(\dv{t}{r})_{\omega_2^\mu }\dd{r} - \int_{r_{\rm e}}^{r_{\rm o1}} \qty(\dv{t}{r})_{\omega_1^\mu }\dd{r} \\ 
		& =\int^{r_{\rm e}}_{r_{\mathrm{c}2}} \frac{\dd{r}}{A(r)\sqrt{1 -b_2^2A(r)/r^2}}+\int_{r_{\mathrm{c}2}}^{r_{\rm o2}} \frac{\dd{r}}{A(r)\sqrt{1 -b_2^2A(r)/r^2}}  - (2\leftrightarrow 1) \ .  
	\end{aligned}
\ee
The same static coordinate time difference can be computed from the geodesic of the observer:
\be \label{dt2}
	t_{\rm o2} - t_{\rm o1} = \int_{r_{\rm o1}}^{r_{\rm o2}} \qty(\dv{t}{r})_{u_{\rm o}^\mu }\dd{r} = \int_{r_{\rm o1}}^{r_{\rm o2}} \frac{E_{\rm o}\dd{r}}{(1 - r_{\rm s}/r - H^2 r^2) \sqrt{E^2_{\rm o} -1 + r_{\rm s}/r + H^2 r^2}} \ .
\ee

\section{Large cosmological distances and the small-angle approximation} \label{s:small_angle}
We work in a regime of large cosmological distances,
\be
H r_{\rm e, o} \lesssim \mathcal{O}(1) \ . 
\ee
Because $ r_{\rm e, o} $ are related to angular diameter distances $ D^{\rm A} $ (see App.~\ref{s:angular_diameter}), this corresponds to $ D^{\rm A}  \sim 1/H \sim $ few Gpc. The causal domain of the static coordinates restricts our analysis to $H r_{\rm e, o}<1$ \footnote{At the same time, in order not to fall too close to the cosmological horizon, we assume $1 - H^2r^2_{\rm e,o} \sim \mathcal{O}(1)$.}. We also assume $ r_{\rm s}/b \ll 1 $ (we suppress the subscript $ _i $ in this section), which is the small-deflection regime. For orientation, a galaxy-scale strong lens with mass $ M \approx 10^{11} M_\odot $ inside the projected Einstein radius has $r_{\rm s}\sim0.01$ pc, while the projected image separation is typically a few kpc. Thus $r_{\rm s}/b$ is of order $10^{-5}$ in the systems that motivate the expansion. More broadly, $r_{\rm s}/b\sim10^{-4}$--$10^{-9}$ characterizes common lensing regimes from weak cosmological lensing to Galactic microlensing, as summarized in App.~\ref{s:x_scales}.

Noting that the impact parameter $b$ is parametrically close to the distance of closest approach, $r_{\rm c}$, we define the small expansion parameter 
\be
	x := \frac{r_{\rm s}}{r_{\rm c}} \ll1\ . 
\ee
It will be useful to express all three roots of Eq.~\eqref{eq:uceq} in terms of $x$. 
The root corresponding to the geodesic turning point is $u_{\rm c}=x$ by definition. Denoting the other two roots by $u_1$ and $u_2$,  we have
\be
	\begin{aligned}
	&u_{\rm c} = x \ , \\ 
	&u_1 = - x +  x^2  - x^3 +2x^4 +\mathcal{O}\qty(x^5) \ , \\  
	&u_2 = 1 - x^2 + x^3 - 2x^4  + \mathcal{O}\qty(x^5) \ .
	\end{aligned}
\ee
We can now expand integrals like those of Eq.~\eqref{b_i} as
\be\label{eq:angle}
	\begin{aligned}
		&\int_{u_{\rm e}}^{u_{\rm c}} \frac{\dd{u}}{\sqrt{x^2(1-x) -u^2(1-u) }} = 
		\frac{1}{\sqrt{u_{\rm c}}} \int_{u_{\rm e}/u_{\rm c}}^{1} \frac{\dd{z}}{\sqrt{(1-z )(z -u_1/u_{\rm c}) (u_2/u_{\rm c} -z) } } \\
		&= \frac{\pi}{2} + x - \frac{u_{\rm e}}{x} + \left(\frac{15\pi}{16} - 1\right)\frac{x^2}{2}  - \frac{u_{\rm e}}{2} +\left(\frac{61}{24}-\frac{15\pi}{32}\right)x^3 - \frac{u_\mathrm{e}^3}{6x^3}- \frac{3}{8}  x u_\mathrm{e} + \mathcal{O}(x^4) \ .
	\end{aligned}
\ee
Below we justify that the scaling $u_{\rm e,o} \sim \mathcal{O}(x^2)$ holds for strong-lensing images at small deflection angle, as long as the images have significant magnification. This scaling organizes the expansion: terms $\sim u_{\rm e,o}$ are counted as $\mathcal{O}(x^2)$. Together with $Hr_{\rm e,o}\lesssim\mathcal{O}(1)$, it implies $Hr_{\rm s}\lesssim\mathcal{O}(x^2)$, or equivalently $Hr_{\rm c}\lesssim\mathcal{O}(x)$, in our setting.
With this expansion, Eq.~\eqref{b_i} becomes
\be
	s(\varphi_{\rm o} - \varphi_{\rm e} + \pi) &=& 2x -\frac{u_{\rm e} + u_{\rm o}}{x}-\frac{u_{\rm e} + u_{\rm o}}{2} + \qty(\frac{15\pi}{16}-1)x^2 + \left(\frac{61}{12} -\frac{15\pi}{16}\right)x^3 \nonumber\\
	&& {} - \frac{3}{8}  x(u_\mathrm{e} + u_\mathrm{o}) - \frac{u_{\rm e}^3+u_{\rm o}^3}{6x^3} + \mathcal{O}(x^4) \ . \label{Angle}
\ee

This is a convenient point to explain the scaling $u_{\rm e,o} \sim \mathcal{O}(x^2)$. To see this, note that an Einstein ring forms when the left hand side of \eqref{Angle} vanishes, namely, $\varphi_{\rm o} - \varphi_{\rm e} + \pi$ is the source position offset $\beta$. At leading order in $x$, the Einstein ring is therefore characterized by $x^2_{\rm E}\approx (u_{\rm e}+u_{\rm o})/2$, so without a parametric hierarchy between $u_{\rm e}$ and $u_{\rm o}$, one has $u_{\rm e,o}\sim \mathcal{O}\left(x_{\rm E}^2\right)$. Since for images with order unity or larger magnification, the impact parameter must obey $x\sim x_{\rm E}$, the assumed scaling is justified.

It is useful to cast the relation between $b$, $\tilde{b}$, and $r_{\rm c}$ in the same language: 
\be
	|b| = |\tilde{b}|\qty(1+\frac{H^2 b^2}{2} + \mathcal{O}(x^4)) = r_{\rm c} \qty(1 + \frac{x}{2} + \frac{3x^2}{8} +\frac{H^2 r_{\rm s}^2}{2x^2} + \mathcal{O}(x^3) ) \ .  
\ee
We thus have
\be
		\frac{r_{\rm s}}{|b|} = x\sqrt{1 - x -H^2 r^2_{\rm c}}= x -\frac{x^2}{2}  - \frac{x^3}{8} - \frac{H^2 r^2_{\rm s}}{2x} +   \mathcal{O}(x^4) \ ,
\ee
which allows us to write the expanded version of Eq.~\eqref{angle_spec}
\be\label{eq:theta}
	\theta = \left(1 + H r_{\rm o}  + \frac{r_{\rm s} }{2 H r^2_{\rm o}}\right)\frac{ |\tilde{b}| }{r_{\rm o}} + (1+ H^3r^3_{\rm o} ) \frac{ |\tilde{b}|^3  }{6 r^3_{\rm o}}  +  \mathcal{O}(x^4) \ .
\ee
Here and below the observer is comoving, $E_{\rm o}=1$ \footnote{We remind the reader that these and subsequent formulas are valid in the regime $H r_{\rm o,e} \gg r_{\rm s}/r_{\rm o,e}$, in particular apparent divergences happening in the limit $H\to 0$ but at finite $r_{\rm s}$ can happen as this limit is outside the regime where the formulas are valid.}.

\subsection{Redshift expansions} \label{s:redshift_expansions}
In order to write the deflection angle equation using the standard notation with angular diameter distances, we need to employ our approximations to the redshift expressions as well, Eq.~\eqref{zs}, \eqref{eq:zl_alpha}.
In the main formulas we use a background lens-redshift parameter $\tilde z_{\rm l}$, which we also called unlensed redshift, defined by the comoving observer relation below. Its relation to a measured lens redshift is discussed in Sec.~\ref{s:exact_redshift}.

The emitter redshift expression Eq.~\eqref{zs} expanded in powers of $x$ yields
\be\label{ze_expanded}
	1 + z_{\rm e} = \frac{1 + H r_{\rm o}}{1-H r_{\rm e}} \qty( 1 - H \frac{r_{\rm e} + r_{\rm o}}{2r_{\rm e}r_{\rm o}}b^2 + \frac{r_{\rm s}}{2 H} \qty( \frac{1}{(1+H r_{\rm o})r_{\rm o}^2} +  \frac{1}{(1-H r_{\rm e})r_{\rm e}^2 }  ) ) + \mathcal{O}(x^3) \ .
\ee

The derivation of the angular diameter distances is discussed in App.~\ref{s:angular_diameter}. We denote by $D_{\rm l}$, $D_{\rm e}$ and $D_{\rm el}$ the observer-lens, observer-emitter and lens-emitter unlensed angular diameter distances\footnote{Unlensed angular diameter distances are defined as the angular diameter distance coming only from the background metric, in our case from the de Sitter expansion alone (and no contribution from the point mass). See App.~\ref{s:angular_diameter} for further details.}, respectively. For our purposes in the main text, we will use the expressions
\be \label{eq:Dd_main}
	D_{\rm l} = \frac{r_{\rm o}}{1+\tilde{z}_{\rm l}}  \ , \ D_{\rm el} = r_{\rm e} \ , \ D_{\rm e} = \frac{r_{\rm e} + r_{\rm o}}{1+\tilde{z}_{\rm l}}  \ ,
\ee
where, as already remarked, $\tilde{z}_{\rm l}$ is a version of the lens redshift which does not suffer from $r_{\rm l} \neq 0$, $E_{\rm l} \neq 1$ corrections. Explicitly,
\be \label{eq:zltilde}
1 + \tilde{z}_{\rm l} = 1+ Hr_{\rm o} + \frac{r_{\rm s}}{2 H r^2_{\rm o}} +\mathcal{O}\qty(x^3) \ .
\ee
The distances in Eqs.~\eqref{eq:Dd_main} are the unlensed angular diameter distances used in the standard lensing notation. They express $r_{\rm o}$ and $r_{\rm e}$ in terms of $D^{\rm A}$, and reduce to the standard angular-diameter-distance expressions in the $r_{\rm s}=0$ limit.

\subsection{The expanded deflection angle equation} \label{s:expanded_deflection}
Using the relation
\be
r_\mathrm{c} =|\tilde{b}| - \frac{r_\mathrm{s} }{2} - \frac{3 r_\mathrm{s}^2 }{8 |\tilde{b}|} - \frac{r_\mathrm{s}^3 }{2 |\tilde{b}|^2} + |\tilde{b}|\mathcal{O}(x^4) \ ,
\ee
and Eq.~\eqref{eq:zltilde} for the lens redshift (which neglects $r_{\rm s}/r_{\rm l}$ corrections), Eq.~\eqref{eq:theta} becomes
\be
\theta = (1 + \tilde{z}_{\rm l}) \frac{ |\tilde{b}| }{r_{\rm o}} + (1+ H^3r^3_{\rm o} ) \frac{ |\tilde{b}|^3  }{6 r^3_{\rm o}}  +  \mathcal{O}(x^5) \ .
\ee
We can then rewrite Eq.~\eqref{Angle}, using the angular diameter distances in Eq.~\eqref{eq:Dd_main}, in the familiar form
\be \label{eq:angle_final}
s\frac{D_{\rm e}}{D_{\rm el}} \beta  =  \frac{D_{\rm e}}{D_{\rm el}} \theta - 2 \frac{r_{\rm s}}{D_{\rm l} \theta} - \frac{15\pi}{16} \qty(\frac{r_{\rm s}}{D_{\rm l} \theta})^2 + \mathcal{O}(x^3) \ .	
\ee
where we defined  $ \beta := - (\varphi_{\rm o} - \varphi_{\rm e} + \pi) D_{\rm el}/D_{\rm e}$. In~Eq.~\eqref{eq:angle_final}, $ \theta $ is positive, and the two image solutions $\vartheta = s\theta$ correspond to $ s=+1 $ and $ s=-1 $. Through $ \mathcal{O}(x^2) $, effects of $ H^2 $ are completely encoded in the unlensed angular diameter distances, in agreement with Ref.~\cite{Butcher:2016yrs}.

Setting $\beta=0$ in Eq.~\eqref{eq:angle_final} gives the Einstein angle
\be \label{eq:einstein_angle}
	\theta_{\rm E} = \sqrt{\frac{2 r_{\rm s} D_{\rm el}}{D_{\rm e} D_{\rm l}}} + \frac{15 \pi }{64} \frac{r_{\rm s}}{D_{\rm l}} + \mathcal{O}(x^3) \ .
\ee
The leading term is the standard Einstein angle for a point mass, and the next-to leading term is the known second-order Schwarzschild correction.

\subsection{Time delays in the small-angle and large-distance approximation} \label{s:approx_time_delay}
We now expand the exact time-delay expressions, Eqs.~\eqref{DeltaT},\eqref{eq:time_delay},\eqref{dt2}.
In our stated approximations, we can write
\be
	\tilde{b}^2 = \frac{r_{\rm c}^3}{r_{\rm c} -r_{\rm s}} = \frac{r_{\rm c}^2}{1 -x}  \ , \ \frac{\tilde{b}}{b} = \sqrt{1 - H^2 \tilde{b}^2} = \sqrt{1 -\frac{ H^2r_{\rm c}^2}{1-x}} = 1 - \frac{1}{2} H^2 r^2_{\rm c} - \frac{1}{2} H^2 r_{\rm c}^2 x +\mathcal{O}(x^3) \ .
\ee
It is also useful to express
\be
	\sqrt{1 - \frac{b^2}{r^2}A(r)} = \frac{b}{\tilde{b}} \sqrt{1 - \frac{r_{\rm c}^2}{r^2(1-x)} \qty(1 - x \frac{r_{\rm c}}{r})  } \ .
\ee
Expanding Eq.~\eqref{eq:time_delay} in $x=r_{\rm s}/r_{\rm c}$, we have\footnote{The $ \mathcal{O}(x^2) $ part of the integrand is involved. Once integrated, part of it contributes at level $r_{\rm c}\mathcal{O}(x^2)=r_{\rm s}\mathcal{O}(x)$ to the time delay. We use it in the expansion leading to Eq.~\eqref{eq:toi}. For completeness, we report it in App.~\ref{s:misc}. Moreover, note the useful relations
	\begin{equation}
		\mathrm{ArcTanh}(x) = \mathrm{ArcTanh}\qty(\frac{1}{x}) - \frac{\iu \pi}{2}  \ , \  \mathrm{ArcTan}(x) = -\mathrm{ArcTan}\qty(\frac{1}{x}) + \frac{\pi}{2} \ .
\end{equation}}
\be
	\begin{aligned} \label{eq:dt_integral}
		\int_{r_{\rm c}}^{r_{\rm e}} &\frac{\dd{r}}{A(r)\sqrt{1 -b^2A(r)/r^2}} \approx  \frac{\tilde{b}}{b}\frac{1}{H\sqrt{1 - H^2 r_{\rm c}^2}} \mathrm{ArcTanh} \frac{H\sqrt{r^2 -r_{\rm c}^2}}{\sqrt{1 - H^2r_{\rm c}^2}} \eval{}_{r_{\rm c}}^{r_{\rm e}} +  \frac{\tilde{b}}{b}\frac{r_{\rm c}x}{2} \Big( \frac{\sqrt{r - r_{\rm c}} (1+H^2r r_{\rm c})}{\sqrt{r + r_{\rm c}} (1-H^2r_{\rm c}^2)(1-H^2r^2)} \\
			&\quad+\frac{2+ Hr_{\rm c} -2H^2r_{\rm c}^2}{(1-H^2r^2_{\rm c})^{3/2}} \mathrm{ArcTanh}\frac{\sqrt{(r-r_{\rm c}) (1+Hr_{\rm c}) }}{\sqrt{(r+r_{\rm c}) (1-Hr_{\rm c}) }} \\
		&\quad+ \frac{2- Hr_{\rm c} -2H^2r_{\rm c}^2}{(1-H^2r^2_{\rm c})^{3/2}} \mathrm{ArcTanh}\frac{\sqrt{(r-r_{\rm c}) (1-Hr_{\rm c}) }}{\sqrt{(r+r_{\rm c}) (1+Hr_{\rm c}) }}  \Big)\eval{}_{r_{\rm c}}^{r_{\rm e}} + r_{\rm c}\mathcal{O}(x^2)  \ .
	\end{aligned}
\ee

We can write the total result for the travel time of image $ i $ as (we now reinstate the subscript $_i$ and define $x_i=r_{\rm s}/r_{\mathrm{c}i}$)
\be \label{eq:dtoi}
	\begin{aligned}
			t_{\mathrm{o}i} &=\frac{\tilde{b}}{b}\frac{1}{H\sqrt{1 - H^2 r_{\mathrm{c}i}^2}} \mathrm{ArcTanh} \frac{H\sqrt{r_{\rm e}^2 -r_{\mathrm{c}i}^2}}{\sqrt{1 - H^2r_{\mathrm{c}i}^2}} +\frac{\tilde{b}}{b}\frac{r_{\rm s}}{2}\Big(  \frac{\sqrt{r_{\rm e} - r_{\mathrm{c}i}} (1+H^2r_{\rm e} r_{\mathrm{c}i})}{\sqrt{r_{\rm e} + r_{\mathrm{c}i}} (1-H^2r_{\mathrm{c}i}^2)(1-H^2r_{\rm e}^2)}	 \\
		&+\frac{2+ Hr_{\mathrm{c}i} -2H^2r_{\mathrm{c}i}^2}{(1-H^2r^2_{\mathrm{c}i})^{3/2}} \mathrm{ArcTanh}\frac{\sqrt{(r_{\rm e}-r_{\mathrm{c}i}) (1+Hr_{\mathrm{c}i}) }}{\sqrt{(r_{\rm e}+r_{\mathrm{c}i}) (1-Hr_{\mathrm{c}i}) }}  \\
			&+ \frac{2- Hr_{\mathrm{c}i} -2H^2r_{\mathrm{c}i}^2}{(1-H^2r^2_{\mathrm{c}i})^{3/2}} \mathrm{ArcTanh}\frac{\sqrt{(r_{\rm e}-r_{\mathrm{c}i}) (1-Hr_{\mathrm{c}i}) }}{\sqrt{(r_{\rm e}+r_{\mathrm{c}i}) (1+Hr_{\mathrm{c}i}) }}  \Big)   + r_{\mathrm{c}i}\mathcal{O}(x_i^2) + (r_{\rm e} \leftrightarrow r_{\mathrm{o}i }) \ .
	\end{aligned}
\ee
Expanding in $ r_{\mathrm{c}i} $ (recalling that $ x_i \sim H r_{\mathrm{c}i} \sim r_{\mathrm{c}i}/r_{\rm e,o} $), the travel time reduces to 
\be\label{eq:toi}
	\begin{aligned}
			t_{\mathrm{o}i} &= \frac{1}{H} \mathrm{ArcTanh} Hr_{\rm e} + \frac{r_{\rm s}}{2} \ln \frac{4 r^2_{\rm e}}{(1-H^2r^2_{\rm e}) r^2_{\rm ci} } + \frac{r_{\rm s}}{2(1-H^2r^2_{\rm e})} - \frac{r_{\rm s} r_{\mathrm{c}i}}{2r_{\rm e}}  \\ 
			& - \frac{r_{\mathrm{c}i}^2}{2r_{\mathrm{e}}}   + \frac{r^2_{\rm s}}{2r_{\rm ci}}\qty( \frac{15\pi}{8}-1) + r_{\rm s} \mathcal{O}(x_i^2) + (r_{\rm e} \leftrightarrow r_{\mathrm{o}i })	 \ .
	\end{aligned}
\ee
Define
\be
y := \frac{r_{\rm o2} - r_{\rm o1}}{r_{\rm o1}} \ .
\ee
In the following image-difference formulas, an unsubscripted $x$ denotes the common parametric size of $x_1$ and $x_2$.
Eq.~\eqref{dt2} implies that $ y/H $ is of the same order as the time delay $ \Delta t $. Since the time delay is of order $r_{\rm s}$ and $Hr_{\rm s}\sim\mathcal{O}(x^2)$, one has $y\sim \mathcal{O}(x^2)$.
Expanding Eq.~\eqref{eq:toi} in $ y $, we obtain
\be \label{dt1}
\Delta t = t_{\mathrm{o}2} - t_{\mathrm{o}1} = \Delta t' + \frac{r_{\rm o1} y}{1- H^2 r_{\rm o1}^2} \qty(1 + \frac{H^2 r^2_{\rm o1}}{1- H^2 r_{\rm o1}^2}y) + r_{\rm s}\mathcal{O}(x^2)\ ,
\ee
where the $y$-independent part is
\be \label{eq:deltatprime}
	\begin{aligned}
		\Delta t' = & 2r_{\rm s} \ln \frac{r_{\rm c1}}{r_{\rm c2}} + \frac{1}{2} (r_{\rm c1}^2 - r_{\rm c2}^2) \qty(\frac{1}{r_{\rm o1}} + \frac{1}{r_{\rm e}} ) + \frac{r_{\rm s}}{2}(r_{\rm c1} - r_{\rm c2})\qty( \frac{1}{r_{\rm e}} + \frac{1}{r_{\rm o1}})  \\
		&  + r^2_{\rm s}\qty(\frac{1}{r_{\rm c2}} - \frac{1}{r_{\rm c1}})\qty( \frac{15\pi}{8}-1) + r_{\rm s}\mathcal{O}(x^2)\ .
	\end{aligned}
\ee

The same coordinate-time difference can be computed using the observer geodesic, Eq.~\eqref{dt2}, which yields 
\begin{equation} \label{eq:dt2}
\Delta t = \frac{y}{H}  \frac{1}{ 1- H^2r_{\rm o1}^2 } + r_{\rm s}\mathcal{O}(x^2) \ .
\end{equation}
Keeping only terms needed for the time delay up to $r_{\rm s}\mathcal{O}(x^2)$, this gives
\be \label{eq:y}
	y = H(1+Hr_{\rm o1})\Delta t' + \mathcal{O}(x^4) \ ,
\ee
where we combined Eq.~\eqref{dt1} and Eq.~\eqref{eq:dt2}.
Using Eq.~\eqref{DeltaT}, the proper-time delay measured by the observer is therefore
\be \label{eq:deltaT_middle}
	\Delta T = (1+Hr_{\rm o1})\Delta t' + r_{\rm s}\mathcal{O}(x^2)
	= (1+\tilde z_{\rm l})\Delta t' + r_{\rm s}\mathcal{O}(x^2)\ ,
\ee
where in the last equality we used Eq.~\eqref{eq:zltilde}. The difference between $1+Hr_{\rm o1}$ and $1+\tilde z_{\rm l}$ starts contributing only at the order included in the remainder.

Using
\be
	\theta_i = (1 + \tilde{z}_{\rm l}) \qty(\frac{|\tilde{b}_i|}{r_{\rm oi}} + \mathcal{O}(x^3)) \implies  r_{\rm ci} = D_{\rm l} \theta_i - \frac{r_{\rm s}}{2 } - \frac{3 r_{\rm s}^2}{8D_{\rm l} \theta_i  }  + D_{\rm l}\mathcal{O}(x^4) \ , \ \frac{1}{r_{\rm o1}} + \frac{1}{r_{\rm e}}  = \frac{D_{\rm e}}{D_{\rm l}D_{\rm el}} \ ,
\ee
we obtain
\be \label{eq:time_delay_final}
	\begin{aligned}
		\Delta T = (1 + \tilde{z}_{\rm l}) \Big( 2r_{\rm s} \ln \frac{\theta_{\rm 1}}{\theta_{\rm 2}} + \frac{1}{2}\frac{D_{\rm e} D_{\rm l}}{D_{\rm el}} (\theta_{\rm 1}^2 - \theta_{\rm 2}^2) 
		+\frac{15\pi r^2_{\rm s}}{8 D_{\rm l}}\qty(\frac{1}{\theta_{\rm 2}} - \frac{1}{\theta_{\rm 1}})  \Big) + r_{\rm s}\mathcal{O}(x^2) \ .
	\end{aligned}
\ee

Equation~\eqref{eq:time_delay_final} is the direct expansion of the exact travel-time integrals. To expose the usual geometrical-plus-Shapiro split, it is useful to use the lens equation once more. Let $\vartheta_i:=s_i\theta_i$ be the signed image position. Up to the same order, Eq.~\eqref{eq:angle_final} lets us rewrite Eq.~\eqref{eq:time_delay_final} as
\be \label{eq:time_delay_Fermat}
	\begin{aligned}
		\Delta T = (1 + \tilde{z}_{\rm l}) \Bigg[&
		\frac{1}{2}\frac{D_{\rm e}D_{\rm l}}{D_{\rm el}}\qty((\vartheta_2-\beta)^2-(\vartheta_1-\beta)^2)
		-2r_{\rm s} \ln \frac{\theta_{\rm 2}}{\theta_{\rm 1}}  \\
		&+\frac{15\pi r^2_{\rm s}}{16 D_{\rm l}}\qty(\frac{1}{\theta_{\rm 2}} - \frac{1}{\theta_{\rm 1}})  \Bigg] + r_{\rm s}\mathcal{O}(x^2) \ .
	\end{aligned}
\ee
The first line is the standard result: a geometrical delay plus a Shapiro delay. The second line is the first correction to this split in the SdS expansion,
\be
	\Delta T_{(1)}=(1+\tilde z_{\rm l})\frac{15\pi r_{\rm s}^2}{16D_{\rm l}}\qty(\frac{1}{\theta_2}-\frac{1}{\theta_1}) \ .
\ee
This correction comes from the Schwarzschild part of the metric.
Since $r_{\mathrm{c} i}\approx D_{\rm l}\theta_i$, this correction is suppressed with respect to the leading time delay by one power of
\be
	x_i=\frac{r_{\rm s}}{r_{\mathrm{c}i}}\approx \frac{r_{\rm s}}{D_{\rm l}\theta_i}\sim \theta_i \ .
\ee
At this order, cosmology enters only through the unlensed angular diameter distances and the redshift prefactor.

\subsubsection{Fermat principle} \label{s:fermat_principle}
Equation~\eqref{eq:time_delay_Fermat} can be read as the difference of a one-image Fermat potential,
\be
	\mathcal{T}(\vartheta;\beta)=(1+\tilde z_{\rm l})\left[
	\frac{1}{2}\frac{D_{\rm e}D_{\rm l}}{D_{\rm el}}(\vartheta-\beta)^2
	-2r_{\rm s}\ln|\vartheta|+\frac{15\pi r_{\rm s}^2}{16D_{\rm l}|\vartheta|}
	\right]\ ,
\ee
evaluated at the two signed image positions, $\Delta T=\mathcal{T}(\vartheta_2;\beta)-\mathcal{T}(\vartheta_1;\beta)+r_{\rm s}\mathcal{O}(x^2)$.
The stationarity statement is that $\mathcal{T}$ should be varied with respect to the image position while holding the source position $\beta$ fixed. Indeed,
\be
	0=\frac{1}{D_{\rm l}(1+\tilde z_{\rm l})}\pdv{\mathcal{T}}{\vartheta}
	=\frac{D_{\rm e}}{D_{\rm el}}(\vartheta-\beta)
	-\frac{2r_{\rm s}}{D_{\rm l}\vartheta}
	-\frac{15\pi r_{\rm s}^2}{16D_{\rm l}^2}\frac{\sgn(\vartheta)}{\vartheta^2}
	+\mathcal{O}(x^3)\ .
\ee
Writing $\vartheta=s\theta$, with $\theta>0$, gives Eq.~\eqref{eq:angle_final}.

\section{Summary} \label{s:summary}
We examined gravitational lensing in the SdS metric using exact geodesic expressions in static coordinates, with a comoving emitter and observer. Expanding in $x=r_{\rm s}/r_{\rm c}\approx r_{\rm s}/(D_{\rm l}\theta)\approx\theta$, we recovered the standard lens equation and found no independent cosmological-constant contribution to the deflection angle through $\mathcal{O}(x^2)$, in agreement with Ref.~\cite{Butcher:2016yrs}.

For the time delay, the standard geometrical-plus-Shapiro formula is the leading term of the same expansion. Eq.~\eqref{eq:time_delay_Fermat} gives the first correction to this split, of relative order $x$. Such correction is just the point mass Schwarzschild correction. While we worked out this exercise in an idealized SdS setup (in particular in a setting which does not handle extended lenses, lens environment, and considers the universe today as $\Lambda$-dominated), we expect that corrections to the Shapiro/geometry split enter at $r_{\rm s}\mathcal{O}(x)$ also in more realistic cosmology and lens scenarios, although dedicated calculations, left for future work, will be needed for confirming such expectation.

\acknowledgments
KB and LT acknowledge the support by the European Research Area (ERA) via the UNDARK project (project number 101159929). LT acknowledges the MICINN through the grant ``DarkMaps'' PID2022-142142NB-I00.

\appendix

\section{Peculiar velocities} \label{s:peculiar}
In the main text we set the emitter and observer to be comoving with the de Sitter expansion, $E_{\rm e}=E_{\rm o}=1$. Here we collect the simple bookkeeping needed to restore small radial peculiar velocities in a few examples. This appendix is not meant as a complete treatment of general peculiar motion. Its purpose is only to indicate how a small radial velocity enters the constants of motion, the observed angle, and the comparison between the two image-observation events.

We begin with the relation between the conserved energy and a local radial peculiar velocity. Consider a comoving observer
\be
	w^\mu=\qty(\frac{1}{A},\sqrt{1-A},0,0)
\ee
and an observer moving with peculiar radial velocity
\be
	u^\mu=\qty(\frac{E}{A},\sqrt{E^2-A},0,0)\ .
\ee
The peculiar velocity $v$ of $u^\mu$ with respect to $w^\mu$ is defined by
\be \label{eq:pec_vel}
	\frac{1}{\sqrt{1-v^2}} := -u^\mu w_\mu
	= \frac{E-\sqrt{1-A(r)}\sqrt{E^2-A(r)}}{A(r)} \ .
\ee
For small $v$ this gives
\be \label{eq:E}
	E = 1 + Hrv + \frac{v^2}{2}+\mathcal{O}(v^3,r_{\rm s}v/r) \ ,
\ee
where the sign convention is that $v>0$ corresponds to motion faster than the comoving outflow.

The same velocity also appears in the observed image angle. In particular, the angle formula Eq.~\eqref{angle_spec} contains the usual aberration of light effect. For example, in the $H=r_{\rm s}=0$ limit one finds
\be
 \cos\theta_i = \frac{\cos(\theta_i(v_{\rm o}=0)) -|v_{\rm o}|}{1-|v_{\rm o}|\cos(\theta_i(v_{\rm o}=0))} \ , \ \cos(\theta_{i}(v_{\rm o}=0)) = \sqrt{1 - \frac{b_i^2}{r^2_{\mathrm{o} i}}} \ .
\ee

The explicit presence of peculiar velocities affects the observed angle equation, Eq.~\eqref{eq:theta},
\begin{equation}
		\theta =  \frac{ |\tilde{b}|  }{ r_{\rm o}}\left[(1 + H r_{\rm o}) \qty(1 + v_{\rm o} +\frac{v^2_{\rm o}}{2})  +\frac{r_{\rm s} }{2 H r^2_{\rm o}} + (1+ H^3r^3_{\rm o} ) \frac{ |\tilde{b}|^2  }{6 r^2_{\rm o}}  
	+ \mathcal{O}(x^3, v^3, v^2 x,v x^2) \right] \ ,
\end{equation}
the redshift equation Eq.~\eqref{ze_expanded}
\begin{align}
\begin{aligned}
1 + z_{\rm e} =& \frac{1 + H r_{\rm o}}{1-H r_{\rm e}}\qty(1 + v_{\rm e} + v_{\rm o} + \frac{(v_{\rm e} + v_{\rm o})^2}{2}) \Bigg( 1 - H \frac{r_{\rm e} + r_{\rm o}}{2r_{\rm e}r_{\rm o}}b^2 \\
&+ \frac{r_{\rm s}}{2 H} \qty( \frac{1}{(1+H r_{\rm o})r_{\rm o}^2} +  \frac{1}{(1-H r_{\rm e})r_{\rm e}^2 } ) \Bigg) 
+ \mathcal{O}(v^3, vr_{\rm s}/r )  \ ,	
\end{aligned}
\end{align}
and the time delay equation Eq.~\eqref{eq:dt2}
\begin{equation}
	\Delta t = \frac{y}{H} \qty( \frac{1}{ 1- H^2r_{\rm o1}^2 }   
	- \frac{v_{\rm o}}{H r_{\rm o}}\qty( 1 +\frac{v_{\rm o}}{2H r_{\rm o}}) +\frac{3v^2_{\rm o}}{2H^2r^2_{\rm o1}}
	)  + \mathcal{O}(v^3, x v^2, x^2) \ ,
\end{equation}
which in turn implies that Eq.~\eqref{eq:y} and Eq.~\eqref{eq:deltaT_middle} read respectively
\begin{align}
	y =& H \Delta t' (1 + H r_{\rm o1})\Big(1 + v_{\rm o} + \frac{v_{\rm o}}{Hr_{\rm o1}} + \frac{v_{\rm o}^2}{Hr_{\rm o1}} (1+Hr_{\rm o1}) 
	+ \mathcal{O}(v^3, v^2 x,x^2 ) \Big)   \ ; \\
	\Delta T =& (1+ Hr_{\rm o1})\qty(1 + v_{\rm o} + \frac{v_{\rm o}^2}{2}) +  r_{\rm s}\mathcal{O}(x^3, v^3, v^2 x,v x^2)  \ .
\end{align}
If we re-define $\tilde{z}$ to contain $v_{\rm o}$ terms, 
\begin{equation}
	1 + \tilde{z}_{\rm l} \to (1 + \tilde{z}_{\rm l} )\qty( 1 + v_{\rm o} +\frac{v^2_{\rm o}}{2} ) + \mathcal{O}(v^3, r_{\rm s} v/r) \ ,
\end{equation}
the main equations, Eq.~\eqref{eq:angle_final} and Eq.~\eqref{eq:time_delay_Fermat} do not change up to $\mathcal{O}(x^3, v^3, v^2 x,v x^2)$ and up to $r_{\rm s}\mathcal{O}(x^2, v^2)$ respectively.


\section{Recovering standard FLRW results for the redshift} \label{s:comoving}
In this appendix, we show how the redshift expressions, Eqs.~\eqref{zs}, \eqref{eq:zl_alpha}, reduce to the expected FLRW results for $ r_{\rm s}  =0$, $ b=0 $ and $E_i=1$. We first write the redshift in the static coordinates used in the main text, and then relate these coordinates to the comoving time $T$.

With these choices, Eq.~\eqref{zs} gives
\be \label{zs_dS}
	1 + z_{\mathrm{e}} = \frac{1+H r_{\mathrm{o}}}{1-Hr_{\rm e}} \ .
\ee

We now connect this expression with the standard FLRW result
\be
1 + z_{\rm e} = \e^{H(T_{\rm o1}-T_{\rm e})} \ ,
\ee
where $ T $ denotes the comoving time at emitter and at the observation events.

The relation between static time and comoving time follows from the massive comoving observer trajectory,
\be
	\dv{r}{t} = \frac{\dot{r}}{\dot{t}} = \pm (1 - H^2 r^2) Hr \ ,
\ee
which integrates to
\be \label{rr}
	\frac{1}{H} \log\frac{r_{\rm o1}\sqrt{1-H^2r^2_{\rm e}}}{r_{\rm e}\sqrt{1-H^2r_{\rm o1}^2}} = t_{\rm o1}-t_{\rm e} \implies \frac{1}{H} \ln\frac{r_{\rm o1}}{r_{\rm e}} = t_{\rm o1}-t_{\rm e} + \frac{1}{2H} \ln \frac{1-H^2 r_{\rm o1}^2}{1-H^2r^2_{\rm e}}  \ .
\ee
Using
\be
	\dv{T}{r} = \frac{1}{Hr} \implies T_{\rm o1} - T_{\rm e} = \frac{1}{H} \ln \frac{r_{\rm o1}}{r_{\rm e}} \ ,
\ee
we can rewrite Eq.~\eqref{rr} as
\be \label{dT}
	T_{\rm e} - T_{\rm o1} = t_{\rm e} - t_{\rm o1} + \frac{1}{2H} \ln\frac{1-H^2 r_{\rm e}^2}{1-H^2r^2_{\rm o1}} \ .
\ee

It remains to use the light-ray propagation time. For a radial massless geodesic,
\be \label{ttlight}
	t_{\rm o1} - t_{\rm e}  = \int_{-r_{\rm e}}^{r_{\rm o1}}  \frac{\dd{r}}{1-H^2r^2} = \frac{1}{2H} \ln \frac{(1+Hr_{\rm e})(1+Hr_{\rm o1})}{(1-Hr_{\rm e})(1-Hr_{\rm o1})} \ .
\ee
Substituting Eq.~\eqref{ttlight} into Eq.~\eqref{rr}, one recovers
\be \label{eq:redshift}
	1 + z = \e^{H(T_{\rm o1} - T_{\rm e})} = \frac{1+Hr_{\rm o1}}{1-Hr_{\rm e}}  \ .
\ee
The corresponding lens-redshift limit follows straightforwardly by setting $ r_{\rm e} =0 $.

\section{Angular diameter distance in a generic metric} \label{s:angular_diameter}
We here go over the calculation of lensed angular diameter distance in the SdS metric, from which we recover the unlensed angular diameter distances used in the main text.

A rigorous definition of angular diameter distance, valid for any metric, reads~\cite{1974ApJ...189..167D,Schneider:1992,Fleury:2014gha}
\be \label{eq:Da_generic}
	\tilde{D}^{\rm A} := \sqrt{\dv{A_{\rm e}}{\Omega_{\rm o}}} \ ,
\ee
where $ \dd{A_{\rm e}} $ is the cross-section area of a light-ray bundle from the emitter and $ \dd{\Omega_{\rm o}} $ is the solid angle which subtends $ \dd{A_{\rm e}} $, as seen by the observer. We use the tilde to distinguish this lensed angular diameter distance from the unlensed angular diameter distance $ D^{\rm A} $ of the main text, which take into account only the de Sitter expansion. 
Lensed angular diameter distances could be observable in principle if the true transverse area $\dd{A_{\rm e}}$ were known. In the lensing applications considered here, this is not the usual situation, and the lens and time-delay equations are therefore expressed in terms of unlensed angular diameter distances connected to observed redshifts. We nevertheless give the calculation of $\tilde{D}^{\rm A}$ in this appendix as an interesting exercise.

To obtain an expression for $\tilde{D}^{\rm A}$, we first need a way to express $\dd{A_{\rm e}}$. We use ray bundles for this purpose. Consider a ray bundle $ x^\mu(\lambda,\zeta_i) $, which represents rays around the source, all ending at the observer point. Here $ \lambda $ is the affine parameter on a ray and the parameters $ \zeta_i $ label the ray. The ray bundle must be differentiable in both $\lambda$ and $\zeta_i$, meaning that events with similar $\zeta_i$ and $\lambda$ must be close. In a general setting, $\zeta_i$ would label three real parameters, e.g. the two angles on the celestial sphere at which the light ray is received at the observer and the phase of the ray~\cite{Schneider:1992}. In the main text, we neglected the motion on the $\phi$ plane, but the angular diameter distance requires it. We therefore discuss the $ \phi $ geodesic motion below.

The geodesic equation for $\phi$ reads
\be\label{eq:ll}
\dv{\lambda} (r^2 \dot{\phi}) = \frac{b^2 \cos\phi}{r^2\sin^2\phi} \ .
\ee
The quantities associated with angular momentum per mass,
\be\label{eq:li}
l_x = -r^2(\dot{\phi} \sin\varphi  + \dot{\varphi}\sin\phi\cos\phi \cos\varphi ) \ , \ l_y = r^2(\dot{\phi}\cos\varphi  -  \dot{\varphi}\sin\phi\cos\phi \sin\varphi) \ ,
\ee
are conserved. Their derivative with respect to $\lambda$ vanishes, as can be verified using Eq.~\eqref{eq:ll} and 
\be
\dot\varphi = \frac{b}{r^2\sin^2\phi} \ .
\ee
Defining $l^2 := l_x^2 + l_y^2 + b^2$ (where $b$ is related to the $z$ component of the angular momentum), we can write the geodesic vector $\omega^\mu$ with the explicit terms of $\phi$ as
\be\label{eq:omegafull}
\omega^\mu = \qty( \frac{1}{A(r)} ,  \pm \sqrt{1 - \frac{l^2 A(r)}{r^2}} , \frac{l_y\cos\varphi  - l_x\sin\varphi }{r^2}, \frac{b}{r^2 \sin^2\phi} ) \ .
\ee
With these constants, Eq.~\eqref{b_i} for $\varphi$ changes to
\be
	\int_{\varphi_{\rm e}}^{\varphi_{\mathrm{o}}} \sin^2\phi \dd{\varphi} = \frac{b}{l} \qty( \int_{u_{\rm e}}^{u_{\mathrm{c}}}  + \int_{u_{\mathrm{o}i}}^{u_{\mathrm{c}}} )  \frac{\dd{u}}{\sqrt{\frac{r^2_{\rm s}}{\tilde{l}^2} -u^2(1-u) }} \ .
\ee
The equations for $t$, Eq.~\eqref{eq:time_delay}, and $r_{\rm c}$, Eq.~\eqref{rc}, are the same except for the substitution $b \leftrightarrow l$. The equation for $\phi$ is also simple. For $ b \neq 0 $,
\be
\frac{\dd{\phi}}{\sin^2\phi} = \qty( l_y \cos\varphi - l_x \sin\varphi ) \frac{\dd\varphi}{b} \ , 
\ee
which implies
\be \label{eq:phi}
\cot \phi_{\rm e } - \cot \phi_{\rm o } = \frac{1}{b}( l_x (\cos\varphi_{\rm e } - \cos\varphi_{\rm o }) - l_y (\sin\varphi_{\rm e } -\sin\varphi_{\rm o })) \ .
\ee
This makes explicit that the geodesic motion happens entirely in a 2D plane.

We now return to the bundle parameters. We take $\zeta_i$ to be the conserved quantities $ b $ and $l_{x,y} $. Then $ \omega^\mu := \dv*{x^\mu}{\lambda} $ is the geodesic vector and $ \xi^\mu := \delta\zeta_i\dv*{x^\mu}{\zeta_i}  $ is the geodesic deviation vector, connected to the distance between neighbouring light rays separated by $ \delta \zeta_i $. We assume the bundle to end at the observer point, so that $\xi^\mu(\lambda=\lambda_{\rm o}) =0$. The following identity will be useful below:
\be
	\dv{\lambda} (\omega^\mu \xi_\mu) = \xi_\mu \underbrace{\omega^\nu \grad_\nu \omega^\mu}_{=0} + \omega^\mu \underbrace{\dv{\xi_\mu}{\lambda}}_{= \delta\zeta_i\dv{\omega_\mu}{\zeta_i}} = \frac{\delta\zeta_i}{2} \dv{\zeta_i} (\omega^\mu \omega_\mu) = 0 \implies \omega^\mu \dot\xi_\mu = 0  \ ,
\ee
where we exploited that $ \omega^\mu $ is a null geodesic vector. 
We next introduce the Sachs screen basis~\cite{1961MAWMN...1....1J,Sachs:1962wk}. Given an observer $ u^\mu $, the screen vectors $ s^\mu_A $, $ A =1,2 $, project on the screen perpendicular to both $ u^\mu $ and $\omega^\mu$.
They satisfy 
\be
g_{\mu\nu}s_A^\mu s_B^\nu = \delta_{AB} \ , \ u_\mu s_A^\mu = \omega_\mu s_A^\mu = 0 \ .
\ee
Defining $ \xi_A := s^\mu_A \xi_\mu $, we can write the Sachs equation, coming from the geodesic deviation equation, as
\be \label{eq:jacobi}
	\dv{^2 \xi_A}{\lambda^2} = \mathcal{R}_{AB} \xi^B \ , \ \mathcal{R}_{AB}  = R_{\mu\nu\alpha\beta} \omega^\nu \omega^\alpha s^\mu_A s^\beta_B \ .
\ee
This equation holds provided that the screen vectors are parallel transported on the plane defined by the projector~\cite{Fleury:2014gha},
\be
	S^{\mu}_\nu = \delta^{AB} g_{\nu \rho} s^\mu_A s^\rho_B \ ,
\ee
namely if
\be \label{eq:Smunu}
	S^\mu_\nu \omega^\rho \grad_\rho s^\nu_A = 0 \ .
\ee 

The transverse components $ \xi_1 $, $ \xi_2 $ determine $ \dd{A_{\rm e}} $. Given the linear nature of Eq.~\eqref{eq:jacobi}, we can write
\be
\xi_A(x_{\rm e}) = C_{AB} \xi_B(x_{\rm o}) + D_{AB} \dot{\xi}_B(x_{\rm o}) \ .
\ee
The relevant factor is $ D_{AB} $, since we assumed $\xi (x_{\rm o}) = 0$. We will show in App.~\ref{s:misc} that the observed angle satisfies $ \theta \sim  \dot{\xi}/\omega_{\rm o}$. Hence
\be \label{eq:DAB}
		\tilde{D}^{\rm A} = \omega_{\rm o}\sqrt{|\det(D)|} \ . 
\ee
The matrix $ D_{AB}$ solves
\be
\dv{^2 D}{\lambda^2} = \mathcal{R} D \ , D(x_{\rm o}) = 0 \ , \  \dot{D}(x_{\rm o}) = \mathbb{I} \ .
\ee

We now specialize this general setup to the SdS metric. Because the point-mass metric is singular at $ r=0 $, the definitions of the lens distance $ \tilde{D}^{\rm A}_{\rm l} $ and the lens-emitter distance $ \tilde{D}^{\rm A}_{\rm el} $ require a prescription for where the lens light is emitted. For a generic light ray, Eq.~\eqref{eq:omega_mu}, and for a generic massive observer, Eq.~\eqref{eq:generic_obs}, using the shorthand $ J $ specified in Eq.~\eqref{eq:zl_alpha},
\be \label{eq:ximu}
	\omega_{\rm o} =\frac{1}{A} \qty(E \mp J(r)\sqrt{1-\frac{b^2}{r^2}A} - \frac{b_u b}{r^2} ) \ .
\ee

In the main lensing setup, the reference geodesic has no motion on $ \phi $. This suggests writing one of the screen vectors as
\be
	s_2^\mu = \qty(0,0,\frac{1}{r},0) \ .
\ee
We first find $ s^\mu_1 $ for a comoving source with $ b_u=0 $. Set $ s^\mu_1 = (a_1, a_2,0,a_3) $. The factors $ a_i $ can be found via
\be
	\begin{aligned}
		&s^\mu_1 u_{\mu} = 0 \implies a_1 = a_2 \frac{\sqrt{E^2-A}}{AE} \ , \\
		&g_{\mu\nu} s_1^\mu s_1^\nu = 1 \implies a_2 = E\sqrt{1-r^2a^2_3}  \ , \\
		&s^\mu_1 \omega_{\mu} = 0 \implies a_3 = \frac{\sqrt{1-r^2 a_3^2}}{bA} \qty(\sqrt{E^2 -A} - E\sqrt{1-\frac{l^2 A}{r^2}})\ .
	\end{aligned}
\ee 
Up to $ \mathcal{O}(x^2) $ in the regime $ r \sim 1/H $, and up to $\mathcal{O}(v)$ (with $v$ peculiar velocities, see Eq.~\eqref{eq:E})
\be\label{eq:sDs}
	\begin{aligned}
	a_1 &= \frac{Hb}{1-Hr}\qty(1 + \frac{v}{Hr}\pm v) + \mathcal{O}\qty(x^3, v^2) \ , \\
	a_2 &=  \frac{b}{r}(1+Hr)(1+ v + vHr)+ \mathcal{O}\qty(x^3, v^2) \ , \\
	a_3 &= \frac{1}{r}\qty(1-\frac{l^2(1+Hr)^2}{2r^2} + \mathcal{O}\qty(x^3, v^2)) \ .
	\end{aligned}
\ee
We show in App.~\ref{s:misc} that $ s_A^\mu $ satisfy the parallel transport condition Eq.~\eqref{eq:Smunu}.

For the lens-distance calculation we also need the screen vectors for a generic emitter with $ b_u\neq 0 $, $ E \neq 1 $ (see Eq.~\eqref{eq:generic_obs}) and for a light ray with $ b=0 $. Without expanding in $x$, these are
\be \label{eq:sDl}
s^\mu_1 = \qty( \frac{b_u}{(E-J(r))r}, \frac{Ab_u}{(E-J(r))r}, 0, \frac{1}{r} ) \ , \ s^\mu_2 = \qty( 0, 0, \frac{1}{r}, 0 ) \ .
\ee

We now compute the matrix $D_{AB}$. Following~\cite{Serbenta:2021tzv}, write
\be
	D_{AB} = \pdv{x^\mu_{\rm E}}{\omega^\nu_{\rm O}} s^A_\mu (x_{\rm E}) s_B^\nu(x_{\rm O}) \ ,
\ee
where the indices $ _{\rm E} $ and $ _{\rm O} $ denote generic emission and observation points, not to be confused with the specific $_{\rm e}$ and $ _{\rm o} $ used in the main lensing setup.
Using Eq.~\eqref{eq:omegafull} and computing derivatives at $l_{x,y}=0$, $\varphi_{\rm o}=0$, $\phi = \pi/2$, the relevant differentials are\footnote{Here, $x^\mu_{\rm e} := x^\mu_{\rm e} (\lambda; x^\nu_{\rm o}, \omega^\rho_{\rm o})$, with $  x^\nu_{\rm o}, \omega^\rho_{\rm o}$ treated as independent variables. In practice, the dependence on $ \omega^\rho_{\rm o}$ of $x^\mu_{\rm e}$ can be translated in the dependence on the constant of motion $l_{x,y}, b$ (for a massless geodesic, there cannot be explicit dependence on $E$ as it can be rescaled away).}
\be
	\dd\omega^r_{\rm o} = -\frac{b A}{\dot{r}_{\rm o}r^2_{\rm o}} \dd{b} \ , \ \dd{\omega^\phi_{\rm o}} = \frac{\dd{l_y}}{r^2_{\rm o}} \ , \ \dd{\omega^\varphi_{\rm o}} = \frac{\dd{b}}{r^2_{\rm o}} \ .
\ee
There are no parameters entering $\omega^t_{\rm o}$, so derivatives with respect to it do not contribute.

We first apply this machinery to $ \tilde{D}^{\rm A}_{\rm l} $. We use a light ray starting at a radius $ r_{\rm lens} > r_{\rm s} $, $ \varphi=0 $, with $ b=0 $ (straight light ray). The observer is comoving, with $ E=E_{\rm o}=1 $ and $ b_u=0 $. Then
\be\label{omegao}
	\omega_{\rm o} = (E_{\rm o}+\sqrt{E^2_{\rm o}-A(r_{\rm o})})  \ .
\ee
The next step is to describe the neighbouring rays in the bundle. Let $x^\mu(\lambda, b,l_{x,y})$ end at $x_{\rm O} $ for a reference $\lambda_{\rm O}$, taken to be zero, and start from $ x^\mu(\lambda_{\rm E}, b,l_{x,y}) $. Here $ \lambda_{\rm E} $ corresponds to the affine parameter of the fiducial light ray, $ x^\mu(\lambda_{\rm E},b=b^{\rm fid},l_{x,y}=l^{\rm fid}_{x,y}) $, with $ b^{\rm fid}=l^{\rm fid}_{x,y}=0 $. The bundle is determined implicitly by
\be
\begin{aligned}
&\int_{\varphi_{\rm O}}^{\varphi(\lambda_{\rm E},b,l_{x,y})} \sin^2\phi \dd{\varphi'}  = \frac{b}{l} \int_{r_{\rm o}}^{r(\lambda,b,l_{x,y})} \frac{\dd{r'}}{r'\sqrt{\frac{{r'}^2}{l^2} - A(r')}} \ ; \\
&t(\lambda_{\rm E},b,l_{x,y}) - t_{\rm O}=	\int_{r(\lambda_{\rm E},b,l_{x,y})}^{r_{\rm O}} \frac{\dd{r'}}{A(r')\sqrt{1 -l^2A(r')/{r'}^2}} \ , \\
&\lambda_{\rm E} = \int_{r_{\rm O}}^{r(\lambda_{\rm E},b,l_{x,y})} \frac{\dd{r'}}{\sqrt{1-l^2A(r')/{r'}^2}} \ . 
\end{aligned}\label{eq:rlambda}
\ee
where $\varphi(\lambda_{\rm E},b,l_{x,y})  $ and $ r(\lambda_{\rm e},b,l_{x,y}) $ are implicitly defined by the angular and radial equations, respectively. 

Differentiating these relations with respect to the bundle parameters gives the ingredients entering $D_{AB}$. With $ \zeta_i $ equal to either $ b, l_x, l_y $,
\be
\begin{aligned}
&\pdv{\zeta_i } \int_{\varphi_{\rm O}}^{\varphi} \sin^2\phi \dd{\varphi'} = \pdv{\varphi}{\zeta_i } \sin^2\phi(\lambda_{\rm E}) 
+  \int_{\varphi_{\rm O}}^{\varphi} 2\sin\phi \cos\phi \pdv{\phi}{\zeta_i } \dd{\varphi'} = \pdv{\varphi}{\zeta_i } \ ,\\
&\pdv{\zeta_i }\cot \phi = \frac{1}{\sin^2\phi} \pdv{\phi}{\zeta_i } = \pdv{\phi}{\zeta_i } \ ,
\end{aligned}
\ee
where the last equalities hold when the expression is evaluated at $ \phi = \pi/2 $. The radial equation similarly gives
\be
\pdv{b}\int_{r_{\rm O}}^{r} \frac{\dd{r'}}{\sqrt{1-l^2A(r')/{r'}^2}} =0 \implies \pdv{r}{b} = - \sqrt{1 -b^2A(r)/r^2} \int_{r_{\rm O}}^{r} \frac{b\dd{r'}}{{r'}^2(1-b^2A(r')/{r'}^2)^{3/2}} \ ,
\ee
Evaluating at $ b=0 $ gives $ \pdv*{r}{b} = 0 $.
At the fiducial ray one also has
\be
\begin{aligned}
&\eval{\pdv{\varphi}{b}}_{b=l_{x,y}=0} =-\int_{r_{\rm E}}^{r_{\rm o}} \frac{\dd{r}}{r^2} = -\qty(\frac{1}{r_{\rm E}} - \frac{1}{r_{\rm O}}) \ ; \\
 &\eval{\pdv{t}{b}}_{b=l_{x,y}=0} = \int_{r}^{r_{\rm O}} \eval{\frac{\dd{r'}A(r)b/{r'}^2}{A(r')\qty(1 -b^2A(r')/{r'}^2)^{3/2}}}_{b=l_{x,y}=0}  
- \frac{1}{A(r)} \eval{\dv{r}{b}}_{b=l_{x,y}=0} = 0\ ; \\
&\eval{\pdv{\phi}{l_i}}_{l_{x,y}=0} = 0 \ . 
\end{aligned}
\ee
The partial derivatives of $ \phi $ at $ b = 0 $ require one extra step. For derivatives with respect to $ l_i $, Eq.~\eqref{eq:li} gives, at $ b=0 $,
\be
\dot{\phi} = \pm\frac{\sqrt{l^2_x + l^2_y}}{r^2} \ ,
\ee
which is analogous to the equation for $ \varphi $. Thus,
\be
\eval{\pdv{\phi}{l_i}}_{l_{x,y}=b=0} =\eval{\pdv{\varphi}{b}}_{b=l_{x,y}=0}  \ .
\ee
It remains to consider the $ \phi $ derivative with respect to $ b $ at $ b=0 $ with $ l_{x,y} =0$. For this case, Eq.~\eqref{eq:li} can be written as
\be
	\begin{pmatrix}
		l_x \\ l_y
	\end{pmatrix} =
\begin{pmatrix}
	\sin\varphi & \cos\varphi \\
	\cos\varphi & -\sin\varphi 
\end{pmatrix} \begin{pmatrix}
r^2\dot{\phi} \\
b\cot\phi
\end{pmatrix} \ .
\ee
Inverting it shows that $ l_{x,y} = 0 $ implies both $ r^2\dot{\phi} = 0 $ and $ b\cot\phi = 0 $. In particular, for $ r \neq 0 $, $ \phi = $ const and there is no dependence on $ b $. Hence $ \pdv{\phi}{b} =0 $.

Combining the preceding derivatives gives
\be
	D_{12} = D_{21} = 0 \ ; \ 	D_{22} = D_{11} = g_{\varphi\varphi}(x_{\rm E}) s_1^{\varphi} (x_{\rm E}) s_1^{\varphi}(x_{\rm O}) \pdv{\varphi(\lambda_{\rm E},b=0,l_{x,y}=0)}{\omega_{\rm O}^\phi} =  -r_{\rm O}\qty(1 - \frac{r_{\rm E}}{r_{\rm O}}) \ .
\ee

For $ \tilde{D}^{\rm A}_{\rm l} $, we substitute $ r_{\rm E} \to r_{\rm lens} $ and $ r_{\rm O} \to r_{\rm o} $. Using Eq.~\eqref{omegao} gives
\be\label{eq:Dl_lensed}
	\tilde{D}^{\rm A}_{\rm l} =  \frac{r_{\rm o}}{E_{\rm o}+\sqrt{E^2_{\rm o} - A(r_{\rm o})}} \qty(1 - \frac{r_{\rm lens}}{r_{\rm o}}) \ .
\ee
If the lens light is emitted from $ r_{\rm lens} \approx r_{\rm c} $, Eq.~\eqref{eq:zltilde} gives
\be 
\tilde{D}^{\rm A}_{\rm l} = \frac{r_{\rm o}}{1+\tilde{z}_{\rm l}} + r_{\rm o}\mathcal{O}(x,v) \ .
\ee

We next consider $ \tilde{D}^{\rm A}_{\rm el}  $. We take a light ray from the source with $ b=0 $ towards a generic non-comoving observer at $ r_{\rm lens} $. 
Using Eq.~\eqref{eq:ximu}, its observed frequency is
\be
	\omega_{\rm o} = \frac{A(r_{\rm lens})}{E + J(r_{\rm lens})} \ ,
\ee
and a calculation analogous to the one for $ \tilde{D}^{\rm A}_{\rm l} $ gives
\be\label{eq:Dls_lensed}
	\tilde{D}^{\rm A}_{\rm el} = r_{\rm e}\frac{ (E+J(r_{\rm lens}))}{A(r_{\rm lens})} \qty(1-\frac{r_{\rm lens}}{r_{\rm e}}) \ .
\ee
By definition, angular diameter distances do not depend on the emitter 4-velocity, as is apparent for $\tilde{D}_{\rm l}^{\rm A}$, but they do depend on the observer 4-velocity. Thus $\tilde{D}_{\rm el}$ depends on the local 4-velocity assigned to the lens. Taking $r_{\rm s}/r_{\rm lens} \sim \mathcal{O}(x)$ and choosing the local observer parameters so that $J(r_{\rm lens})=0$, $E = 1 + \mathcal{O}(x^2)$, as in the derivation of Eq.~\eqref{eq:zltilde}, gives
\be
	\tilde{D}^{\rm A}_{\rm el} = r_{\rm e}(1+\mathcal{O}(x)) \ .
\ee

Finally, for $ \tilde{D}^{\rm A}_{\rm e}  $, all derivatives are evaluated at $ l_{x,y}=0 $ and generic $ b $. The mixed derivatives vanish:
\be
\pdv{r}{l_i} = \pdv{t}{l_i} = \pdv{\varphi}{l_i} = 
	\pdv{\phi}{b} = 0 \ , 
\ee
so $ D_{12} = D_{21} =0 $. 
The remaining derivatives can be counted in the small-$x$ expansion. Using Eq.~\eqref{eq:toi}, and substituting $r_{\rm e} \to r(\lambda,b, l_{x,y})$, the $ \pdv*{t}{b} = s\pdv*{t}{r_{\rm c}} + \mathcal{O}(x^2)$ term yields only $\mathcal{O}(x)$ terms,
\be
\pdv{t}{b} = -s\qty(2\frac{r_{\rm s}}{r_{\rm c}} + r_{\rm c} \frac{r_{\rm e} +r_{\rm o}}{r_{\rm e} r_{\rm o} }) + \mathcal{O}(x^2) \ .
\ee
Because of the structure of the $ s^\mu_1 $ terms, time derivatives start entering only at second order\footnote{Schematically, one obtains $ \pdv{t}{\omega^\nu_{\rm O}} s^A_t (x_{\rm E}) s_B^\nu(x_{\rm O}) \propto r_{\rm o} \mathcal{O}(x) \pdv{t}{b} $.}. The same applies for $ \pdv*{r}{b} $ (using the equivalent of Eq.~\eqref{eq:rlambda} for the light ray emitter-observer and deriving it with respect to $ b $). 
The only non-zero term at $ \mathcal{O}(x) $ for $ D_{11} $ arises from $ \pdv*{\varphi}{b} $. Using Eq.~\eqref{Angle} with $ \varphi_{\rm o} =0 $, we obtain
\be
D_{11} =r^2_{\rm e} s_1^\varphi(x_{\rm e}) s_1^\varphi(x_{\rm o}) r_{\rm o}^2 \pdv{\varphi}{b} =  r_{\rm o}\qty(\frac{r_{\rm e} + r_{\rm o}}{r_{\rm o}} + 2r_{\rm e}\frac{ r_{\rm s}}{\tilde{b}^2} + \frac{15}{8} \pi \frac{r_{\rm e} r^2_{\rm s}}{\tilde{b}^3} ) + r_{\rm o} \mathcal{O}(x^2) \ .
\ee
The second diagonal component follows from Eq.~\eqref{eq:phi}. Using $\sin(-\varphi_{\rm e}) = \sin(-\varphi_{\rm e} +\pi) \approx -\varphi_{\rm e} +\pi $ to connect it to Eq.~\eqref{Angle}, we find
\be
D_{22} = r_{\rm o}^2 \pdv{\phi}{l_y} = -\frac{r_{\rm o}^2}{b}\sin\varphi_{\rm e} \approx \frac{r_{\rm o}^2}{b}(-\varphi_{\rm e} +\pi) \ .
\ee
Combining these terms, and using $  \tilde{b} \approx b $ at this order, gives
\be  \label{eq:Ds_lensed}
	\begin{aligned}
	\tilde{D}^{\rm A}_{\rm e} &= \frac{r_{\rm o}}{1+\tilde{z}_{\rm l}} \sqrt{\qty(\frac{r_{\rm e} + r_{\rm o}}{r_{\rm o}} + 2\frac{r_{\rm e} r_{\rm s}}{\tilde{b}^2}+ \frac{15}{8} \pi \frac{r_{\rm e} r^2_{\rm s}}{\tilde{b}^3} ) \qty( \frac{r_{\rm e} + r_{\rm o}}{r_{\rm o}} - 2\frac{r_{\rm e} r_{\rm s}}{\tilde{b}^2}- \frac{15}{16} \pi \frac{r_{\rm e} r^2_{\rm s}}{\tilde{b}^3}  )} + 	r_{\rm e} \mathcal{O}(x^2) \\ 
	&=  \frac{r_{\rm o} +r_{\rm e} }{1+\tilde{z}_{\rm l}} \sqrt{1 - 4\frac{r^2_{\rm e}r^2_{\rm o} r^2_{\rm s}}{(r_{\rm o }+r_{\rm e})^2\tilde{b}^4} }  + 	r_{\rm e} \mathcal{O}(x) \ .
	\end{aligned}
\ee
The result is cumbersome, and does not simplify the lens equation relative to the unlensed angular diameter distances used in the main text. 

For $\beta = 0$, upon substituting $b \approx \tilde{b}$ with $\theta_{\rm E}$ of Eq.~\eqref{eq:einstein_angle}, we have $ \tilde{D}^{\rm A}_{\rm e} = 0 + r_{\rm e}\mathcal{O}(x^2)$. This is expected because the point is a caustic, or equivalently a conjugate point to the vertex at $x_{\mathrm{o}}$ in the language of Jacobi fields. The same conclusion follows from the fact that angular diameter distances are proportional to the inverse magnification. At the caustic, the magnification is formally infinite in the geometric optics limit.

Finally, with $ r_{\rm s} = 0 $ and in the $ b\to 0 $ limit for $ D_{\rm e} $, Eqs.~\eqref{eq:Dl_lensed}, \eqref{eq:Dls_lensed}, and \eqref{eq:Ds_lensed} reduce to the unlensed angular diameter distance expressions used in the main text,
\be \label{eq:Dd}
D_{\rm l} = \frac{r_{\rm o}}{1+\tilde{z}_{\rm l}}  \ , \ 	D_{\rm el} = r_{\rm e} \ , \ D_{\rm e} = \frac{r_{\rm e} + r_{\rm o}}{1+\tilde{z}_{\rm l}}  \ . 
\ee

Angular diameter distances are insensitive to the emitter peculiar velocity $v_{\rm e}$, whereas the sensitivity on $v_{\rm o}$ would manifest itself in extra contributions to $\tilde{z}_{\rm l}$. The asymmetry between $(r_{\rm e},v_{\rm e})$ and $(r_{\rm o},v_{\rm o})$ follows from the fact that angular diameter distances depend on the 4-velocity of the observer, not of the emitter. By contrast, the luminosity distance is sensitive to $v_{\rm e}$ through the Etherington distance-duality relation~\cite{etherington1933lx}, $D_{\rm e}^{\rm L} = (1+z_{\rm e})^2 D_{\rm e}^{\rm A}$.

\section{Astrophysical size of the expansion parameter} \label{s:x_scales}
This appendix gives order-of-magnitude values of the expansion parameter in a few common lensing settings. The expansion parameter used in the main text is
\be
x=\frac{r_{\rm s}}{r_{\rm c}}\approx \frac{2GM(<b)}{b}\ ,
\ee
where $b\approx r_{\rm c}$ is the physical impact parameter.  In an Einstein-ring configuration this can also be written as
\be
x_{\rm E} \approx \frac{1}{2}\theta_{\rm E}\frac{D_{\rm s}}{D_{\rm ls}}\ ,
\ee
with the usual angular-diameter distances. This relation is useful because typical image separations are often quoted directly.

For galaxy-scale time-delay cosmography, the lenses are usually massive early-type galaxies producing image separations of order a few arcseconds, with typical delays of days to months~\cite{Birrer:2019stz2254}. Taking $\theta_{\rm E}\sim1''$ and $D_{\rm s}/D_{\rm ls}\sim1.5$--$3$ gives
\be
x_{\rm E}\sim 4\times10^{-6}\text{--}8\times10^{-6}\ .
\ee
Equivalently, a mass of order $10^{11}M_\odot$ inside a projected radius of a few kpc gives $x\sim10^{-5}$. This is the regime most directly relevant for the expansion in the main text.

Weak cosmological lensing is not usually described by a single point-mass lens, but the same compactness estimate gives useful orientation. Cosmic shear produces coherent shape distortions at the percent level~\cite{Zhang:2016nww017}. For galaxy-galaxy weak lensing, a characteristic halo mass $M\sim10^{12}h^{-1}M_\odot$ over projected radii $\sim100$--$200h^{-1}\,{\rm kpc}$~\cite{Brainerd:1996jp} gives
\be
x\sim 5\times10^{-7}\text{--}10^{-6}\ .
\ee
For cluster weak lensing, massive clusters have masses $M\sim10^{14}$--$10^{15}h^{-1}M_\odot$~\cite{Umetsu:2020aaz}. A concrete weak-lensing measurement of the Coma cluster gives $M_{200}\approx1.9\times10^{15}h^{-1}M_\odot$ and $r_{200}\approx2h^{-1}{\rm Mpc}$~\cite{Kubo:2007coma}. At radii of order $1$--$3\,{\rm Mpc}$ this corresponds roughly to
\be
x\sim {\rm few}\times10^{-5}\text{--}10^{-4}\ .
\ee

For Galactic microlensing by stellar lenses, the natural scale is the physical Einstein radius. Standard microlensing configurations have $M\sim0.5$--$1M_\odot$, relative parallaxes $\pi_{\rm rel}\sim0.03$--$0.4\,{\rm mas}$, and angular Einstein radii $\theta_{\rm E}\sim0.3$--$1.7\,{\rm mas}$~\cite{Cassan:2016stw372}. For a bulge-like example with $D_{\rm l}=4\,{\rm kpc}$ and $D_{\rm s}=8\,{\rm kpc}$,
\be
R_{\rm E}\approx 4.0\,{\rm AU}\sqrt{\frac{M}{M_\odot}}\ ,\qquad
x_{\rm E}\approx 4.9\times10^{-9}\sqrt{\frac{M}{M_\odot}}\ .
\ee
Thus stellar microlensing typically probes $x\sim10^{-9}$--$10^{-8}$.

These estimates put the expansion parameter of this paper in context: the common lensing applications range from $x\sim10^{-4}$ in massive cluster weak lensing to $x\sim10^{-9}$ in stellar microlensing, with galaxy-scale time-delay lenses near $x\sim10^{-5}$.

\section{Miscellaneous} \label{s:misc}
This appendix collects two short technical checks used in App.~\ref{s:angular_diameter}, and the $r_{\rm c}\mathcal{O}(x^2)$ term of Eq.~\eqref{eq:dt_integral}.

\paragraph{Parallel transport of screen vectors.}
As a sanity check, we verify that $s^\mu_1$ as written in Eq.~\eqref{eq:sDs} obeys Eq.~\eqref{eq:Smunu}. Keeping the terms relevant at the order of the calculation,
\be
	\begin{aligned}
		S^t_t &= -\frac{H^2b^2}{1-Hr} (1+Hr) \ , \ S^t_r = \frac{Hb^2}{r(1-Hr)^2}  \ , S^t_\varphi = \frac{Hbr}{1-Hr}  \ , \\
		S^r_t &= -\frac{Hb^2}{r} (1+Hr)^2 \ , \ S^r_r = \frac{b^2}{r^2}\frac{1+Hr}{1-Hr}  \ ,\ S^r_\varphi = b(1+Hr)  \ , \\
		S^\varphi_t &= \frac{Hb}{r} (1+Hr) \ , \ S^\varphi_r = \frac{b}{r^2}\frac{1}{1-Hr}  \ ,\ S^\varphi_\varphi = 1  \ ,
	\end{aligned}
\ee  
we have the Christoffel symbols~\cite{Muller:2009bw}
\be
	\Gamma^t_{rt} = -\Gamma^r_{rr} = -\frac{H^2r}{1-H^2r^2} \ , \ \Gamma^r_{tt} = -(1-H^2r^2) H^2r \ , \ \Gamma^r_{r\varphi} = \frac{1}{r} \ , \ \Gamma^r_{\varphi\varphi} = -(1-H^2r^2) r \ .
\ee
Together with
\be
\begin{aligned}
	k^\nu \grad_\nu s^t_1 = k^r \partial_r s^t_1 + k^r \Gamma^t_{rt} k^t + k^t \Gamma^t_{tr} k^r = 0 \ , \  
	k^\nu \grad_\nu s^r_1 = -\frac{2b}{r^2}(1-H^2r^2) \ , \ 
	k^\nu \grad_\nu s^\varphi_1 = -\frac{2b^2}{r^4}(1+Hr) \ ,
\end{aligned}
\ee
these expressions verify Eq.~\eqref{eq:Smunu}. 
The vector $ s_2^\mu $ satisfies the same transport condition because $ \Gamma^\phi_{r\phi} =1/r $. A similar calculation applies to the $ s^\mu_A $ of Eq.~\eqref{eq:sDl}. The useful identities are
\be
A\omega^{\rho} \grad_\rho s^t_1 = \omega^{\rho} \grad_\rho s^r_1 \ , \ S^t_t = -A S^t_r \ , \ S^r_t = -A S^r_r \ , \ S^\varphi_t = -A S^\varphi_r \ .
\ee
The remaining components, such as $ \omega^{\rho} \grad_\rho s^\varphi_1 $ and $ \omega^{\rho} \grad_\rho s^\phi_2 $, vanish in the same way.

\paragraph{Justification of $ \omega_{\rm o} $ factor in the angular diameter distance definition.}
We justify the statement $\theta \sim \dot{\xi}/\omega_{\rm o}$, and hence the $\omega_{\rm o}$ factor in the definition Eq.~\eqref{eq:DAB}.  
For simplicity we neglect the motion on $\phi$, so there is no dependence on $l_{x,y}$. Applying Eq.~\eqref{angle_spec} to the two neighbouring light rays at the observer location, $ \omega^\mu(\lambda_{\rm o},b) $ and $ \omega^\mu(\lambda_{\rm o}, b + \delta b) $, and expanding up to second order in $ \delta b $, gives
\be
	1 - \cos\theta \approx \frac{\delta\theta^2}{2} = - \frac{\omega^\mu(\lambda_{\rm o},b) \omega_\mu(\lambda_{\rm o},b+\delta b)}{u^\mu \omega_\mu(\lambda_{\rm o},b) u^\nu \omega_\nu(\lambda_{\rm o},b+\delta b) }
	\approx - \frac{\overbrace{\omega^\mu\dot\xi_\mu}^{=0} \delta b +  \frac{\omega^\mu(\lambda_{\rm o},b)}{2} \pdv{^2 \omega_\mu}{b^2} \delta b^2}{\omega^2_{\rm o}} \ ,
\ee
where we defined $ \omega_{\rm o} = -g_{\mu\nu} u^\mu_{\rm o} \omega^\nu(\lambda_{\rm o},b) $. Noticing
\be
	\omega_\mu \pdv{\dot\xi^\mu}{b} = \pdv{b} \overbrace{(\omega^\mu \dot\xi_\mu)}^{=0} - \dot\xi^\mu\dot\xi_\mu \ ,
\ee
one obtains
\be
	\dd\theta_{\rm o} = \frac{\dd b}{\omega_{\rm o}} \sqrt{\dot\xi^\mu\dot\xi_\mu} \ ,
\ee
which agrees with the expected $ \theta \sim \dot{\xi}/\omega_{\rm o} $. 

\paragraph{$r_{\rm c}\mathcal{O}(x^2)$ term of Eq.~\eqref{eq:dt_integral}.}
It reads
\begin{align}
\begin{aligned}
	\frac{1}{16} x^2 \Bigg(
	&- \frac{
		2 \sqrt{r^2 - r_{\mathrm{c}}^2} \Bigg(
		\frac{H^2 \big(7 + 2 H^2 (r - 3 r_{\mathrm{c}}) r_{\mathrm{c}}\big)}{-1 + H^2 r^2}
		+ \frac{4 - 6 H^2 r_{\mathrm{c}}^2}{r r_{\mathrm{c}} + r_{\mathrm{c}}^2}
		+ \frac{1 - H^2 r_{\mathrm{c}}^2}{(r + r_{\mathrm{c}})^2}
		+ \frac{2 H^2 (-1 + H^2 r_{\mathrm{c}}^2)}{(-1 + H^2 r^2)^2}
		\Bigg) r_{\mathrm{s}}^2
	}{
		(-1 + H^2 r_{\mathrm{c}}^2)^2
	}
	\\[6pt]
	&+ \frac{30 r_{\mathrm{s}}^2}{r_{\mathrm{c}}} \qty(\arctan\!\left(\frac{\sqrt{r^2 - r_{\mathrm{c}}^2}}{r_{\mathrm{c}}}\right) - \frac{\pi}{2})
	\\[6pt]
	&+ \frac{
		H \Big(
		4 H^4 r_{\mathrm{c}}^6 - 19 r_{\mathrm{s}}^2 - 4 H r_{\mathrm{c}} r_{\mathrm{s}}^2 + 4 H^3 r_{\mathrm{c}}^3 r_{\mathrm{s}}^2 
		- H^2 r_{\mathrm{c}}^4 (8 + 15 H^2 r_{\mathrm{s}}^2) + r_{\mathrm{c}}^2 (4 + 31 H^2 r_{\mathrm{s}}^2)
		\Big)
		\ln\!\big(H(-1 + Hr)\big)
	}{
		(1 - H^2 r_{\mathrm{c}}^2)^{5/2}
	}
	\\[6pt]
	&+ \frac{
		H \Big(
		4 H^4 r_{\mathrm{c}}^6 - 19 r_{\mathrm{s}}^2 + 4 H r_{\mathrm{c}} r_{\mathrm{s}}^2 - 4 H^3 r_{\mathrm{c}}^3 r_{\mathrm{s}}^2 
		- H^2 r_{\mathrm{c}}^4 (8 + 15 H^2 r_{\mathrm{s}}^2) + r_{\mathrm{c}}^2 (4 + 31 H^2 r_{\mathrm{s}}^2)
		\Big)
		\ln\!\big(H(1 + Hr)\big)
	}{
		(1 - H^2 r_{\mathrm{c}}^2)^{5/2}
	}
	\\[6pt]
	&- \frac{
		H \Big(
		4 H^4 r_{\mathrm{c}}^6 - 19 r_{\mathrm{s}}^2 - 4 H r_{\mathrm{c}} r_{\mathrm{s}}^2 + 4 H^3 r_{\mathrm{c}}^3 r_{\mathrm{s}}^2 
		- H^2 r_{\mathrm{c}}^4 (8 + 15 H^2 r_{\mathrm{s}}^2) + r_{\mathrm{c}}^2 (4 + 31 H^2 r_{\mathrm{s}}^2)
		\Big)
	}{
		(1 - H^2 r_{\mathrm{c}}^2)^{5/2}
	}
	\\
	& \quad \quad \times \ln\!\big(-r + H r_{\mathrm{c}}^2 - \sqrt{r^2 - r_{\mathrm{c}}^2}\sqrt{1 - H^2 r_{\mathrm{c}}^2}\big) \\[6pt]
	&- \frac{
		H \Big(
		4 H^4 r_{\mathrm{c}}^6 - 19 r_{\mathrm{s}}^2 + 4 H r_{\mathrm{c}} r_{\mathrm{s}}^2 - 4 H^3 r_{\mathrm{c}}^3 r_{\mathrm{s}}^2 
		- H^2 r_{\mathrm{c}}^4 (8 + 15 H^2 r_{\mathrm{s}}^2) + r_{\mathrm{c}}^2 (4 + 31 H^2 r_{\mathrm{s}}^2)
		\Big)
	}{
		(1 - H^2 r_{\mathrm{c}}^2)^{5/2}
	} 
	\\
	&\quad \quad  \times \ln\!\big(r + H r_{\mathrm{c}}^2 - \sqrt{r^2 - r_{\mathrm{c}}^2}\sqrt{1 - H^2 r_{\mathrm{c}}^2}\big) \eval{\Bigg)}^{r_{\rm e}}_{r_{\rm c}} \ .
\end{aligned}
\end{align}

\bibliography{ref}

\providecommand{\href}[2]{#2}\begingroup\raggedright\begin{thebibliography}{10}

\bibitem{Schneider:1992}
P.~{Schneider}, J.~{Ehlers}, and E.~E. {Falco},
  \href{http://dx.doi.org/10.1007/978-3-662-03758-4}{{\em {Gravitational
  Lenses}}}.
\newblock 1992.

\bibitem{Seitz:1994xf}
S.~Seitz, P.~Schneider, and J.~Ehlers, ``{Light propagation in arbitrary
  space-times and the gravitational lens approximation},''
  \href{http://dx.doi.org/10.1088/0264-9381/11/9/016}{{\em Class. Quant. Grav.}
  {\bfseries 11} (1994) 2345--2374},
  \href{http://arxiv.org/abs/astro-ph/9403056}{{\ttfamily
  arXiv:astro-ph/9403056}}.

\bibitem{Bartelmann:2010fz}
M.~Bartelmann, ``{Gravitational Lensing},''
  \href{http://dx.doi.org/10.1088/0264-9381/27/23/233001}{{\em Class. Quant.
  Grav.} {\bfseries 27} (2010) 233001},
  \href{http://arxiv.org/abs/1010.3829}{{\ttfamily arXiv:1010.3829
  [astro-ph.CO]}}.

\bibitem{Suyu:2016qxx}
S.~H. Suyu {\em et~al.}, ``{H0LiCOW ? I. H0 Lenses in COSMOGRAIL's Wellspring:
  program overview},'' \href{http://dx.doi.org/10.1093/mnras/stx483}{{\em Mon.
  Not. Roy. Astron. Soc.} {\bfseries 468} no.~3, (2017) 2590--2604},
\href{http://arxiv.org/abs/1607.00017}{{\ttfamily arXiv:1607.00017
  [astro-ph.CO]}}.

\bibitem{Millon:2019slk}
M.~Millon {\em et~al.}, ``{TDCOSMO. I. An exploration of systematic
  uncertainties in the inference of $H_0$ from time-delay cosmography},''
  \href{http://dx.doi.org/10.1051/0004-6361/201937351}{{\em Astron. Astrophys.}
  {\bfseries 639} (2020) A101},
  \href{http://arxiv.org/abs/1912.08027}{{\ttfamily arXiv:1912.08027
  [astro-ph.CO]}}.

\bibitem{Sasaki:1993tu}
M.~Sasaki, ``{Cosmological gravitational lens equation: Its validity and
  limitation},'' \href{http://dx.doi.org/10.1143/PTP.90.753}{{\em Prog. Theor.
  Phys.} {\bfseries 90} (1993) 753--781}.

\bibitem{Wucknitz:2020spz}
O.~Wucknitz, L.~G. Spitler, and U.~L. Pen, ``{Cosmology with gravitationally
  lensed repeating Fast Radio Bursts},''
  \href{http://dx.doi.org/10.1051/0004-6361/202038248}{{\em Astron. Astrophys.}
  {\bfseries 645} (2021) A44},
  \href{http://arxiv.org/abs/2004.11643}{{\ttfamily arXiv:2004.11643
  [astro-ph.CO]}}.

\bibitem{Lake:1977ui}
K.~Lake and R.~C. Roeder, ``{Effects of a Nonvanishing Cosmological Constant on
  the Spherically Symmetric Vacuum Manifold},''
  \href{http://dx.doi.org/10.1103/PhysRevD.15.3513}{{\em Phys. Rev. D}
  {\bfseries 15} (1977) 3513--3519}.

\bibitem{Perlick:2004tq}
V.~Perlick, ``{Gravitational lensing from a spacetime perspective},''
  \href{http://dx.doi.org/10.12942/lrr-2004-9}{{\em Living Rev. Rel.}
  {\bfseries 7} (2004) 9}.

\bibitem{1918AnP...361..401K}
F.~{Kottler}, ``{{\"U}ber die physikalischen Grundlagen der Einsteinschen
  Gravitationstheorie},''
  \href{http://dx.doi.org/10.1002/andp.19183611402}{{\em Annalen der Physik}
  {\bfseries 361} no.~14, (Jan., 1918) 401--462}.

\bibitem{Piattella:2016nzt}
O.~F. Piattella, ``{On the effect of the cosmological expansion on the
  gravitational lensing by a point mass},''
  \href{http://dx.doi.org/10.3390/universe2040025}{{\em Universe} {\bfseries 2}
  no.~4, (2016) 25}, \href{http://arxiv.org/abs/1609.00270}{{\ttfamily
  arXiv:1609.00270 [gr-qc]}}.

\bibitem{Rindler:2007zz}
W.~Rindler and M.~Ishak, ``{Contribution of the cosmological constant to the
  relativistic bending of light revisited},''
  \href{http://dx.doi.org/10.1103/PhysRevD.76.043006}{{\em Phys. Rev. D}
  {\bfseries 76} (2007) 043006},
  \href{http://arxiv.org/abs/0709.2948}{{\ttfamily arXiv:0709.2948
  [astro-ph]}}.

\bibitem{Ishak:2010zh}
M.~Ishak and W.~Rindler, ``{The Relevance of the Cosmological Constant for
  Lensing},'' \href{http://dx.doi.org/10.1007/s10714-010-0973-9}{{\em Gen. Rel.
  Grav.} {\bfseries 42} (2010) 2247--2268},
  \href{http://arxiv.org/abs/1006.0014}{{\ttfamily arXiv:1006.0014
  [astro-ph.CO]}}.

\bibitem{Aghili:2014aga}
M.~E. Aghili, B.~Bolen, and L.~Bombelli, ``{Effect of accelerated global
  expansion on the bending of light},''
  \href{http://dx.doi.org/10.1007/s10714-016-2175-6}{{\em Gen. Rel. Grav.}
  {\bfseries 49} no.~1, (2017) 10},
  \href{http://arxiv.org/abs/1408.0786}{{\ttfamily arXiv:1408.0786 [gr-qc]}}.

\bibitem{Hu:2021yzn}
L.~Hu, A.~Heavens, and D.~Bacon, ``{Light bending by the cosmological
  constant},'' \href{http://dx.doi.org/10.1088/1475-7516/2022/02/009}{{\em
  JCAP} {\bfseries 02} no.~02, (2022) 009},
  \href{http://arxiv.org/abs/2109.09785}{{\ttfamily arXiv:2109.09785
  [astro-ph.CO]}}.

\bibitem{Sereno:2008kk}
M.~Sereno, ``{The role of Lambda in the cosmological lens equation},''
  \href{http://dx.doi.org/10.1103/PhysRevLett.102.021301}{{\em Phys. Rev.
  Lett.} {\bfseries 102} (2009) 021301},
  \href{http://arxiv.org/abs/0807.5123}{{\ttfamily arXiv:0807.5123
  [astro-ph]}}.

\bibitem{Arakida:2011ty}
H.~Arakida and M.~Kasai, ``{Effect of the cosmological constant on the bending
  of light and the cosmological lens equation},''
  \href{http://dx.doi.org/10.1103/PhysRevD.85.023006}{{\em Phys. Rev. D}
  {\bfseries 85} (2012) 023006},
  \href{http://arxiv.org/abs/1110.6735}{{\ttfamily arXiv:1110.6735 [gr-qc]}}.

\bibitem{Butcher:2016yrs}
L.~M. Butcher, ``{No practical lensing by Lambda: Deflection of light in the
  Schwarzschild\textendash{}de Sitter spacetime},''
  \href{http://dx.doi.org/10.1103/PhysRevD.94.083011}{{\em Phys. Rev. D}
  {\bfseries 94} no.~8, (2016) 083011},
  \href{http://arxiv.org/abs/1602.02751}{{\ttfamily arXiv:1602.02751 [gr-qc]}}.

\bibitem{1975ApJ...195L..11C}
J.~H. {Cooke} and R.~{Kantowski}, ``{Time Delay for Multiply Imaged Quasars},''
  \href{http://dx.doi.org/10.1086/181697}{{\em \apjl} {\bfseries 195} (Jan.,
  1975) L11}.

\bibitem{Refsdal:1964nw}
S.~Refsdal, ``{On the possibility of determining Hubble's parameter and the
  masses of galaxies from the gravitational lens effect},'' {\em Mon. Not. Roy.
  Astron. Soc.} {\bfseries 128} (1964) 307.

\bibitem{Kagramanova:2006ax}
V.~Kagramanova, J.~Kunz, and C.~Lammerzahl, ``{Solar system effects in
  Schwarzschild-de Sitter spacetime},''
  \href{http://dx.doi.org/10.1016/j.physletb.2006.01.069}{{\em Phys. Lett. B}
  {\bfseries 634} (2006) 465--470},
  \href{http://arxiv.org/abs/gr-qc/0602002}{{\ttfamily arXiv:gr-qc/0602002}}.

\bibitem{Schucker:2008jv}
T.~Schucker and N.~Zaimen, ``{Cosmological constant and time delay},''
  \href{http://dx.doi.org/10.1051/0004-6361:200809449}{{\em Astron. Astrophys.}
  {\bfseries 484} (2008) 103}, \href{http://arxiv.org/abs/0801.3776}{{\ttfamily
  arXiv:0801.3776 [astro-ph]}}.

\bibitem{McVittie:1933zz}
G.~C. McVittie, ``{The mass-particle in an expanding universe},''
  \href{http://dx.doi.org/10.1093/mnras/93.5.325}{{\em Mon. Not. Roy. Astron.
  Soc.} {\bfseries 93} (1933) 325--339}.

\bibitem{Kaloper:2010ec}
N.~Kaloper, M.~Kleban, and D.~Martin, ``{McVittie's Legacy: Black Holes in an
  Expanding Universe},''
  \href{http://dx.doi.org/10.1103/PhysRevD.81.104044}{{\em Phys. Rev. D}
  {\bfseries 81} (2010) 104044},
  \href{http://arxiv.org/abs/1003.4777}{{\ttfamily arXiv:1003.4777 [hep-th]}}.

\bibitem{Park:2008ih}
M.~Park, ``{Rigorous Approach to the Gravitational Lensing},''
  \href{http://dx.doi.org/10.1103/PhysRevD.78.023014}{{\em Phys. Rev. D}
  {\bfseries 78} (2008) 023014},
  \href{http://arxiv.org/abs/0804.4331}{{\ttfamily arXiv:0804.4331
  [astro-ph]}}.

\bibitem{Bessa:2022sdh}
P.~Bessa and O.~F. Piattella, ``{Gravitational lensing in a universe with
  matter and a cosmological constant},''
  \href{http://dx.doi.org/10.1103/PhysRevD.106.123513}{{\em Phys. Rev. D}
  {\bfseries 106} no.~12, (2022) 123513},
  \href{http://arxiv.org/abs/2209.04063}{{\ttfamily arXiv:2209.04063
  [astro-ph.CO]}}.

\bibitem{Arakida:2011st}
H.~Arakida, ``{Application of Time Transfer Function to McVittie Spacetime:
  Gravitational Time Delay and Secular Increase in Astronomical Unit},''
  \href{http://dx.doi.org/10.1007/s10714-011-1170-1}{{\em Gen. Rel. Grav.}
  {\bfseries 43} (2011) 2127--2139},
  \href{http://arxiv.org/abs/1103.2569}{{\ttfamily arXiv:1103.2569 [gr-qc]}}.

\bibitem{Einstein:1945id}
A.~Einstein and E.~G. Straus, ``{The influence of the expansion of space on the
  gravitation fields surrounding the individual stars},''
  \href{http://dx.doi.org/10.1103/RevModPhys.17.120}{{\em Rev. Mod. Phys.}
  {\bfseries 17} (1945) 120--124}.

\bibitem{1954ZPhy..137..595S}
E.~{Sch{\"u}cking}, ``{Das Schwarzschildsche Linienelement und die Expansion
  des Weltalls.},'' \href{http://dx.doi.org/10.1007/BF01375011}{{\em
  Zeitschrift fur Physik} {\bfseries 137} no.~5, (Oct., 1954) 595--603}.

\bibitem{Schucker:2008bc}
T.~Schucker, ``{Strong lensing in the Einstein-Straus solution},''
  \href{http://dx.doi.org/10.1007/s10714-008-0731-4}{{\em Gen. Rel. Grav.}
  {\bfseries 41} (2009) 1595--1610},
  \href{http://arxiv.org/abs/0807.0380}{{\ttfamily arXiv:0807.0380
  [astro-ph]}}.

\bibitem{Boudjemaa:2011ab}
K.-E. Boudjemaa, M.~Guenouche, and S.~R. Zouzou, ``{Time delay in the
  Einstein-Straus solution},''
  \href{http://dx.doi.org/10.1007/s10714-011-1152-3}{{\em Gen. Rel. Grav.}
  {\bfseries 43} (2011) 1707--1731},
  \href{http://arxiv.org/abs/1006.0080}{{\ttfamily arXiv:1006.0080
  [astro-ph.CO]}}.

\bibitem{Chen:2010gi}
B.~Chen, R.~Kantowski, and X.~Dai, ``{Time Delay in Swiss Cheese Gravitational
  Lensing},'' \href{http://dx.doi.org/10.1103/PhysRevD.82.043005}{{\em Phys.
  Rev. D} {\bfseries 82} (2010) 043005},
  \href{http://arxiv.org/abs/1006.3500}{{\ttfamily arXiv:1006.3500
  [astro-ph.CO]}}.

\bibitem{Alchera:2017sjt}
N.~Alchera, M.~Bonici, and N.~Maggiore, ``{Towards a new proposal for the time
  delay in gravitational lensing},''
  \href{http://dx.doi.org/10.3390/sym9100202}{{\em Symmetry} {\bfseries 9}
  no.~10, (2017) 202}, \href{http://arxiv.org/abs/1709.09055}{{\ttfamily
  arXiv:1709.09055 [astro-ph.CO]}}.

\bibitem{Alchera:2018ijo}
N.~Alchera, M.~Bonici, R.~Cardinale, A.~Domi, N.~Maggiore, C.~Righi, and
  S.~Tosi, ``{Analysis of the angular dependence of time delay in gravitational
  lensing},'' \href{http://dx.doi.org/10.3390/sym10070246}{{\em Symmetry}
  {\bfseries 10} no.~7, (2018) 246},
  \href{http://arxiv.org/abs/1804.03111}{{\ttfamily arXiv:1804.03111
  [astro-ph.CO]}}.

\bibitem{Perlick:2003vg}
V.~Perlick, ``{On the Exact gravitational lens equation in spherically
  symmetric and static space-times},''
  \href{http://dx.doi.org/10.1103/PhysRevD.69.064017}{{\em Phys. Rev. D}
  {\bfseries 69} (2004) 064017},
  \href{http://arxiv.org/abs/gr-qc/0307072}{{\ttfamily arXiv:gr-qc/0307072}}.

\bibitem{Perlick:2015ncm}
V.~Perlick, ``{Gravitational lensing beyond the weak-field approximation},''
  \href{http://dx.doi.org/10.1063/1.4861947}{{\em AIP Conf. Proc.} {\bfseries
  1577} no.~1, (2015) 94--111}.

\bibitem{1974ApJ...189..167D}
C.~C. {Dyer} and R.~C. {Roeder}, ``{Observations in Locally Inhomogeneous
  Cosmological Models},'' \href{http://dx.doi.org/10.1086/152784}{{\em \apj}
  {\bfseries 189} (Apr., 1974) 167--176}.

\bibitem{Fleury:2014gha}
P.~Fleury, ``{Swiss-cheese models and the Dyer-Roeder approximation},''
  \href{http://dx.doi.org/10.1088/1475-7516/2014/06/054}{{\em JCAP} {\bfseries
  06} (2014) 054}, \href{http://arxiv.org/abs/1402.3123}{{\ttfamily
  arXiv:1402.3123 [astro-ph.CO]}}.

\bibitem{1961MAWMN...1....1J}
P.~{Jordan}, J.~{Ehlers}, and R.~K. {Sachs}, ``{Beitr{\"a}ge zur Theorie der
  reinen Gravitationsstrahlung},'' {\em Mainz Akademie Wissenschaften
  Mathematisch Naturwissenschaftliche Klasse} {\bfseries 1} (Aug., 1961) 1.

\bibitem{Sachs:1962wk}
R.~K. Sachs, ``{Gravitational waves in general relativity. 8. Waves in
  asymptotically flat space-times},''
  \href{http://dx.doi.org/10.1098/rspa.1962.0206}{{\em Proc. Roy. Soc. Lond. A}
  {\bfseries 270} (1962) 103--126}.

\bibitem{Serbenta:2021tzv}
J.~Serbenta and M.~Korzy{\'n}ski, ``{Bilocal geodesic operators in static
  spherically-symmetric spacetimes},''
  \href{http://dx.doi.org/10.1088/1361-6382/ac79f3}{{\em Class. Quant. Grav.}
  {\bfseries 39} no.~15, (2022) 155002},
  \href{http://arxiv.org/abs/2111.02880}{{\ttfamily arXiv:2111.02880 [gr-qc]}}.

\bibitem{etherington1933lx}
I.~Etherington, ``Lx. on the definition of distance in general relativity,''
  {\em The London, Edinburgh, and Dublin Philosophical Magazine and Journal of
  Science} {\bfseries 15} no.~100, (1933) 761--773.

\bibitem{Birrer:2019stz2254}
S.~Birrer and T.~Treu, ``{Astrometric requirements for strong lensing
  time-delay cosmography},''
  \href{http://dx.doi.org/10.1093/mnras/stz2254}{{\em Mon. Not. Roy. Astron.
  Soc.} {\bfseries 489} no.~2, (2019) 2097--2103}.

\bibitem{Zhang:2016nww017}
J.~Zhang, ``{The quest for an accurate measurement of cosmic shear},''
  \href{http://dx.doi.org/10.1093/nsr/nww017}{{\em Natl. Sci. Rev.} {\bfseries
  3} no.~2, (2016) 159--160}.

\bibitem{Brainerd:1996jp}
T.~G. Brainerd, R.~D. Blandford, and I.~Smail, ``{Weak Gravitational Lensing by
  Galaxies},'' \href{http://dx.doi.org/10.1086/177537}{{\em Astrophys. J.}
  {\bfseries 466} (1996) 623}.

\bibitem{Umetsu:2020aaz}
K.~Umetsu, ``{Cluster--galaxy weak lensing},''
  \href{http://dx.doi.org/10.1007/s00159-020-00129-w}{{\em Astron. Astrophys.
  Rev.} {\bfseries 28} (2020) 7}.

\bibitem{Kubo:2007coma}
J.~M. Kubo, A.~Stebbins, J.~Annis, I.~P. Dell'Antonio, H.~Lin, H.~Khiabanian,
  and J.~A. Frieman, ``{The Mass of the Coma Cluster from Weak Lensing in the
  Sloan Digital Sky Survey},'' \href{http://dx.doi.org/10.1086/523101}{{\em
  Astrophys. J.} {\bfseries 671} no.~2, (2007) 1466--1470},
  \href{http://arxiv.org/abs/0709.0506}{{\ttfamily arXiv:0709.0506
  [astro-ph]}}.

\bibitem{Cassan:2016stw372}
A.~Cassan and C.~Ranc, ``{Interferometric observation of microlensing
  events},'' \href{http://dx.doi.org/10.1093/mnras/stw372}{{\em Mon. Not. Roy.
  Astron. Soc.} {\bfseries 458} no.~2, (2016) 2074--2079}.

\bibitem{Muller:2009bw}
T.~Muller and F.~Grave, ``{Catalogue of Spacetimes},''
  \href{http://arxiv.org/abs/0904.4184}{{\ttfamily arXiv:0904.4184 [gr-qc]}}.

\end{thebibliography}\endgroup
\bibliographystyle{utphys}

\end{document}